\documentclass[aps,prd,twocolumn,showpacs,superscriptaddress,groupedaddress,eqsecnum,nofootinbib,floatfix]{revtex4-2}

\usepackage{amsmath}
\usepackage{amsfonts}
\usepackage{amssymb}	
\usepackage{graphicx}
\usepackage{bm}
\usepackage{color}
\usepackage{commath}

\usepackage{graphicx}
\usepackage{dcolumn}

\usepackage{color}
\usepackage{xcolor}
\usepackage{subfigure}

\definecolor{linkcolor}{rgb}{0.0,0.3,0.5}
\usepackage[unicode,colorlinks=true,citecolor=linkcolor,linkcolor=linkcolor,urlcolor=linkcolor]{hyperref}

\usepackage{cleveref}

\graphicspath{{../figures/}}

\newcommand{\tj}[1]{{\textcolor{green! 45! black}{\texttt{TJ: #1}} }}

\allowdisplaybreaks
\begin{document}
	
	\title{Radiation reaction force for scalar-tensor theories in effective-one-body formalism}.
	
	\author{Tamanna \surname{Jain}$^{1,2}$}
	\email{tj317@cam.ac.uk}
	\author{Piero \surname{Rettegno}$^{3}$}
	\email{piero.rettegno@to.infn.it}
	\affiliation{${}^1$ LPENS, Département de physique, Ecole normale supérieure, Université PSL, Sorbonne Université, Université Paris Cité, CNRS, 75005 Paris.}
	\affiliation{${}^2$Department of Applied Mathematics and Theoretical Physics,University of Cambridge,Wilberforce Road CB3 0WA Cambridge, United Kingdom.}%
	\affiliation{${}^3$Dipartimento di Fisica e Geologia, Università di Perugia,
	INFN Sezione di Perugia, Via Pascoli, I-06123 Perugia, Italy}
	\date{\today}
	\begin{abstract}
	Whilst most of the binary configurations in modified theories of gravity are studied under quasi-circular orbit limit, eccentricity effects could play a significant role in future gravitational wave detections. We derive the gravitational radiation-reaction force for nonspinning eccentric orbits within the effective-one-body (EOB) description for the massless scalar-tensor theories up to 1.5 post-Newtonian (PN) order. The effects in such theories start at $1/c^3$  and interestingly, the 1.5PN order effect is due to the radiation reaction square effects which is a conservative effect. The results derived here can be implemented in the quasi-circular EOB-based waveform models to construct waveform templates for generic orbit binaries within massless scalar-tensor theories of gravity.

	\end{abstract} 
	
	\maketitle
	\section{Introduction}
	The detection of gravitational wave events by the LIGO collaboration \cite{LIGOScientific:2016aoc} opened the possibility of studying the dynamics in the strong-gravity regime. 
	Future gravitational wave events measured by the current \cite{Reitze:2019iox,Punturo:2010zz,Amaro-Seoane:2018gbb} and next generation of detectors, such as the Einstein telescope \cite{Maggiore:2019uih} and the Cosmic Explorer \cite{Evans:2021gyd}, will open the possibility of testing General Relativity (GR) in strong-gravity regime by constraining the parameters of alternative gravity theories.
	
The simplest extension of GR to consider scalar fields non-minimally coupled to gravity, called scalar-tensor theories,  have been extensively studied in the literature both observationally~\cite{Damour:1996ke, Freire:2012mg, Shao:2017gwu, Kramer:2021jcw, Zhao:2022vig, Gautam:2022cpb}, numerically~\cite{Healy:2011ef, Barausse:2012da, Berti:2013gfa, Shibata:2013pra, Palenzuela:2013hsa, Taniguchi:2014fqa}, and theoretically~\cite{Damour:1992we, Damour:1993hw, Mirshekari:2013vb, Zaglauer:1992bp, Will:2014kxa, Lang:2013fna, Lang:2014osa, Bernard:2018hta, Bernard:2018ivi, Brax:2021qqo, Shiralilou:2021mfl,
 Khalil:2022sii, Bernard:2022noq, Bernard:2023eul,AbhishekChowdhuri:2022ora},  especially within the post-Newtonian (PN) formalism.  The energy flux in scalar-tensor theories is know up to 1.5PN order for generic (non-circular orbits without tides and spins) orbits \cite{Bernard:2022noq} and the angular momentum flux for inspiralling compact objects for generic orbits up to 1.5PN order was recently computed in Refs.~\cite{Jain:2024lie}. 

In addition to the post-Newtonian description, there has also been recent progress in constructing the effective one body (EOB) Hamiltonian from the two-body PN dynamics in scalar-tensor theories for non-spinning binary~\cite{Julie:2017pkb,Julie:2017ucp,Jain:2022nxs,Jain:2023fvt,Julie:2022qux} with leading order (LO) tidal effects~\cite{Jain:2022nxs}. 

The EOB formalism \cite{Buonanno:1998gg,Buonanno:2000ef} constructed by combining the information of the post-Newtonian (PN) theory and numerical relativity (NR) waveforms improves the waveform generation and hence is crucial to produce semi-analytical accurate waveform templates to match filter against the data observed in the gravitational wave detectors. 
In the context of scalar-tensor theories, EOB-based models have always employed quasi-circular radiation reaction force and waveform. The aim of this paper is to extend the quasi-circular radiation-reaction force for the scalar-tensor theories to eccentric orbits up to 1.5PN order.

The paper is organized as follows. In Sec. \ref{Sec:Two}, we give a brief overview the scalar-tensor theories followed by the brief overview of the EOB description and the radiation reaction force in Sec.~\ref{Sec:Three}. In Sec.~\ref{Sec:Four}, we first derive a transformation between ADM, Harmonic and EOB coordinate systems up to 2PN order, explore the arbitrariness (due to gauge choice) in the reactive terms at 1.5PN order using the method developed by Iyer and Will in Ref.~\cite{Iyer:1993xi, Iyer:1995rn} and then finally present the energy and angular momentum flux in EOB coordinates up to 2PN order. Finally, in Sec.~\ref{sec:Five}, we first discuss the approach used in this work and then compute the radiation-reaction force up to 0.5PN order, followed by a discussion in Sec.~\ref{sec:Six}.

In this work, we use the conventions of Refs.~\cite{Jain:2022nxs,Jain:2023fvt,Julie:2022qux, Jain:2024lie} to define the body-dependent ST parameters (see Appendix A of Ref.~\cite{Jain:2024lie} for details). Unless other otherwise specified, we will also consider geometric units, such as $G=c=1$.

	\section{Brief Overview: Scalar-tensor}
	\label{Sec:Two}
	
	In this work, we consider the massless scalar-tensor theories for which the action is described by the coupling of the usual Einstein-Hilbert action with the scalar field $\varphi$ as
\begin{align}
S&=\frac{c^4}{16 \pi G}\int d^4x \sqrt{-g}(R-2g^{\mu \nu} \partial_\mu \varphi \partial_\nu \varphi)\nonumber\\
  &\qquad\qquad\qquad\qquad+S_m[\Psi, {\mathcal{A}(\varphi)}^2g_{\mu\nu}]~.
\end{align}
Here, $g_{\mu\nu}$ is the Einstein metric, $R$ is the Ricci scalar, $\Psi$ collectively denotes the matter fields, $g \equiv \det(g_{\mu\nu})$ and $G$ is the bare Newton's constant. 
The above action is expressed in the Einstein frame, which is obtained by a conformal transformation of the physical frame (Jordan frame) metric $\tilde{g}_{\mu \nu}$ as
\begin{equation}
\tilde{g}_{\mu\nu}={\mathcal{A}(\varphi)}^2 g_{\mu\nu} \ ,
\end{equation}
 to simplify the calculations, where $\mathcal{A}(\varphi)$ is called the coupling function. The scalar field couples minimally to Einstein frame metric and non-minimally to the matter fields, however it is non-minimally coupled to the metric in the Jordan frame.
The scalar-tensor theory is uniquely fixed when the function ${\mathcal{A}(\varphi)}$ is defined, 
General Relativity is recovered when ${\mathcal{A}(\varphi)}= cst$. 
We adopt the conventions and notations of Damour and Esposito-Farese (DEF, hereafter)~\cite{Damour:1992we,Damour:1995kt}. 

The field equations for the scalar-tensor theories in Einstein frame are derived in \cite{Damour:1992we} and the coupling between the matter and the scalar field in the equations of motion is measured by the parameter
\begin{equation}
\label{alpha-consta}
\alpha(\varphi)=\frac{\partial \ln \mathcal{A}}{\partial \varphi}~.
 \end{equation}

The effective gravitational constant in scalar-tensor theories depends on the scalar-field so that the compact object size and its internal gravity vary with the scalar field.  
Therefore, as suggested by~\cite{1975ApJ...196L..59E}, the compact, self-gravitating objects in scalar-tensor theories can be considered as point particles described by the mass function $m_J(\phi)$, dependent on the value of scalar field at the position of particle.
The matter action is then given by
\begin{equation}
S_{m}=-\sum_{J=A,B}\int \sqrt{-g_{\mu \nu}\frac{dx^{\mu}}{d\lambda}\frac{d x^{\nu}}{d\lambda}} {m}_{J}(\varphi)~,
\end{equation}
where ${m}_J(\varphi)$ is the Einstein frame mass of body $J$, and $\lambda$
is an affine parameter.
Since $ \tilde{g}_{\mu\nu}={\mathcal{A}(\varphi)}^2 g_{\mu\nu}$, the Jordan frame mass is then defined as
\begin{equation}
\tilde{m}_J(\varphi)=\frac{{m}_J(\varphi)}{\mathcal{A}(\varphi)}~.
\end{equation}
The dimensionless body-dependent parameters that describe the scalar field effect in the Einstein-frame up to the 2PN order are defined using the Einstein frame mass function $m_J(\varphi)$ following Refs.~\cite{Damour:1992we,Damour:1995kt} \ i.e.
\begin{align}
\alpha_J&=\frac{d\ln m_J(\varphi)}{d\varphi}, \nonumber\\
\beta_J&=\frac{d\alpha_J}{d\varphi}~.
\nonumber\\
\beta'_J&=\frac{d\beta_J}{d\varphi}~.
\end{align} 
The energy flux in Refs.~\cite{Bernard:2022noq,Lange:2017wki,Lange:2018pyp} is expressed in Jordan frame equivalents of the above parameters. 
As we work in DEF conventions, we will make use of the relations given in Table I of Ref.~\cite{Jain:2022nxs} to perform the conversion of Jordan frame parameters to DEF ones.

\section{Brief Overview: EOB \& radiation reaction force}
\label{Sec:Three}
In this section, before proceeding to the computations of the radiation-reaction force in ST theory, we briefly review the EOB formalism proposed in Refs.~\cite{Buonanno:1998gg,Buonanno:2000ef} as a way to extend the validity of PN results beyond the weak-field and slow-motion regime by resumming the PN results. The three main features of the EOB approach are: description of the conservative (Hamiltonian) dynamics of a two-body system, expression for the radiation-reaction effects, and the description of GWs emitted by the coalescence of a compact binary.

The EOB approach maps the relative motion of two bodies with masses $m_A$ and $m_B$ into the motion of an effective particle of mass $\mu = m_A m_B / M$, where $M = m_A + m_B$, moving in an external metric. Throughout this paper, we describe the motion of the effective particle with rescaled polar coordinates $\left(r,\,\theta,\,\phi\right)$.

The conservative EOB dynamics is completely described by the following effective metric up to 2PN order
\begin{equation}
	\label{eq:EOBconstr}
ds^2_{\rm eff} = -A(r)c^2 dt^2_{\rm eff} + B(r) dr^2+ r^2\left(d\theta^2+ \sin(\theta)^2d\phi^2\right)
\end{equation}
where $A(r)$ and $B(r)$ are the two EOB potentials, $t_{\rm eff}$ is the coordinate time of the effective EOB metric, and $r$ is the radial separation in EOB coordinates. 
In this work, we specialise to nonspinning systems and equatorial motions, i.e. set $\theta = {\pi}/{2}$. 
The generalisation of the EOB formalism beyond 2PN order is attained by introducing the non-geodesic contributions to the effective metric (see, e.g.,  Ref.~\cite{Damour:2000gg} for details). 

The metric above, Eq.~\eqref{eq:EOBconstr}, can be used to obtain an EOB Hamiltonian, hereafter denoted by $H$, describing the conservative dynamics of generic binary orbits, both bound and unbound. 
The EOB Hamiltonian in the scalar-tensor theories for non-spinning binary is known up to 3PN order\cite{Julie:2017pkb,Julie:2017ucp,Jain:2022nxs,Jain:2023fvt,Julie:2022qux}. 
However, the current EOB realization of the radiation reaction force and the emitted waveform in scalar-tensor theories is limited to the case of quasi-circular orbits. 
The main aim of our study is to extend the radiation reaction force along general orbits for nonspinning compact objects.

In GR, different approaches have been employed to study the gravitational radiation reaction. In this work, we will follow the \textit{balance} approach developed in Refs.~\cite{Iyer:1993xi,Iyer:1995rn}, i.e.~we relate the energy and angular momentum losses of the system to the corresponding gravitational radiation at infinity.
This approach was already used in Refs.~\cite{Bini:2012ji, Khalil:2021txt} to derive the radiation reaction force for GR within the EOB-formalism.

For nonprecessing binaries, when the motion is planar, Hamilton's equations read 
\begin{align}
	\label{eq:Heq}
	\dot{r} &= \frac{\partial {H}}{\partial p_r}~,\quad \dot{p} _r= -\frac{\partial {H}}{\partial r} + \mathcal{F}_r~,\nonumber\\
	\dot{\phi} &= \frac{\partial {H}}{\partial p_{\phi}}~,\quad \dot{p} _{\phi}= -\frac{\partial {H}}{\partial \phi} + \mathcal{F}_{\phi}\,,
\end{align}
where `dot' denotes differentiation with respect to time and radiation reaction forces ($\mathcal{F}_r,\mathcal{F}_\varphi$)~\cite{Buonanno:2000ef,Buonanno:2005xu} have been added to the conservative equations. 

The force components $ \mathcal{F}_{\phi}$ and $ \mathcal{F}_r$ can be determined by computing the variation in the energy
\begin{align}
E_{\rm {system}} = H (r,p_r, p_{\phi}) - (m_A+ m_B) c^2~,
\end{align}
and the angular momentum
\begin{align}
J_{\rm {system}}  = p_{\phi}~,
\end{align}
of the binary system using Hamilton's equations, Eq.~\eqref{eq:Heq} above, and comparing them with the corresponding energy $\Phi_E$ and angular momentum $\Phi_J$ fluxes emitted in the form of gravitational waves at infinity. 
However, we can not simply equate the energy and angular momentum losses of the system to $-\Phi_E$ and $-\Phi_J$, respectively, as the energy and angular momentum of the system have additional terms called the Schott terms~\cite{Schott}, say $E_{\rm Schott}$ and $J_{\rm Schott}$, due to the system's interaction with the radiation field.
 
Using Hamilton's equations, and remembering $H$ is independent of $\phi$, we obtain that the time derivatives of the energy and angular momentum (radiation losses) of the system are
 \begin{align}
 \label{eq:EandJsys}
 \dot{E}_{\rm {system}}& =\dot{r}\, \mathcal{F}_{r}+\dot{\phi}\,\mathcal{F}_{\phi}~,\nonumber\\
 \dot{J}_{\rm {system}}&=\mathcal{F}_{\phi}~.
 \end{align}
 
The balance equation after including the Schott terms read
\begin{align}
\label{eq:EandJbalancelaw}
 \dot{E}_{\rm {system}}+ \dot{E}_{\rm {Schott}}+\Phi_E&=0~,\nonumber\\
  \dot{J}_{\rm {system}}+ \dot{J}_{\rm {Schott}}+\Phi_J&=0~.
\end{align}
Combining Eqs.~\eqref{eq:EandJbalancelaw} and~\eqref{eq:EandJsys}, we obtain the equations to solve for the two components of the force are
\begin{align}
\label{eq:rrforceecc}
\mathcal{F}_{r}\dot{r}&=-\mathcal{F}_{\phi}\dot{\phi}- \dot{E}_{\rm {Schott}}-\Phi_E~,\nonumber\\
\mathcal{F}_{\phi}&=- \dot{J}_{\rm {Schott}}-\Phi_J~.
\end{align}

The far zone energy and angular momentum fluxes are gauge invariant but the radiation reaction forces $\mathcal{F}_{r}$ and $\mathcal{F}_{\phi}$, and the Schott terms could be gauge dependent quantities. The gauge freedom in radiation reaction force at the leading order in GR, i.e.~2.5PN order, was first discussed by Refs.~\cite{Iyer:1993xi, Iyer:1995rn} where they found that this gauge freedom is related to the freedom to change the coordinate system. Following this, Ref.~\cite{Bini:2012ji} showed that the gauge freedom of radiation reaction term at 2.5PN in GR is also related to gauge freedom in defining Schott terms, i.e., we can consider a vanishing ${J}_{\rm {Schott}}$. 

However,  due to the presence of a non-vanishing dipole in scalar-tensor theories, in addition to the quadrupolar radiation, there is also a scalar dipolar radiation in scalar-tensor theories starting at a lower PN order. In Sec.~\ref{sec:htoadm}, we study the presence (or lack of) gauge freedom at 1.5PN order.

	\section{Coordinate system Transformations}
	\label{Sec:Four}
	In Refs.~\cite{Bernard:2022noq, Jain:2024lie}, the angular momentum flux and the energy flux for scalar-tensor theory are derived in harmonic coordinate system. As the main aim of this paper is to derive the radiation-reaction within the EOB formalism, we transform the fluxes into EOB coordinates. 
	To do so, we use (i) the transformation between ADM and EOB coordinates derived in Refs.~\cite{Julie:2017pkb, Julie:2017ucp, Jain:2022nxs} and (ii) the transformation between ADM and harmonic coordinates derived in Refs.~\cite{Julie:2017pkb, Jain:2022nxs}. 
	
	
	Here, we first define the  family of fundamental scalars associated to each coordinate system
	\begin{align}
		v_h^2 &= (\textbf{v}_h\cdot\textbf{v}_h),~~\dot{r}_h^2 = (\textbf{n}_h\cdot\textbf{v}_h)^2,~~ \frac{1}{r_h}~,\nonumber\\
		v_a^2 &= (\textbf{v}_a\cdot\textbf{v}_a) ,~~p_a^2 = (\textbf{p}_a\cdot\textbf{p}_a),~~\dot{r}_a^2 = (\textbf{n}_a\cdot\textbf{v}_a)^2,\nonumber \\
		p_{r, a}^2 &= (\textbf{n}_a\cdot\textbf{p}_a)^2,~~ \frac{1}{r_a}~,\nonumber\\
		\textrm{and}\nonumber\\
		p_e^2 &= (\textbf{p}_e\cdot\textbf{p}_e),~~p_{r, e}^2 = (\textbf{n}_e\cdot\textbf{p}_e)^2,~~ \frac{1}{r_e}~,
	\end{align}
	where the subscript `$h$', `$a$' and `$e$' indicates harmonic, ADM and EOB coordinate system respectively. The bold faced letters are used to denote (three-dimensional) spatial vectors.
	We will also use `tilde' to define dimensionless variables, i.e.,
	\begin{equation}
	\tilde{\textbf{p}}_X = \frac{\textbf{p}_X}{\mu}~,\hspace{0.6cm}\tilde{p}_{r, X}^2 = \left(\frac{p_{r, X}}{\mu}\right)^2~,
	\end{equation}
	for $X = \{h,a,e\}$.
	
	\subsubsection{ADM to harmonic}
	\label{sec:htoadm}
	
As the acceleration dependence in the Lagrangian starts at 2PN order, the transformation between ADM and harmonic coordinate system also starts at 2PN order. The explicit 2PN coordinate transformation is derived in Ref.~\cite{Julie:2017pkb} and the explicit 3PN coordinate transformation is derived in Ref.~\cite{Jain:2022nxs}. The transformation up to 2PN order is,
\begin{widetext}
	\begin{align}
		\textbf{r}_h &= \textbf{r}_a + \frac{1}{c^4}\left\{\textbf{r}_a\left[\frac{G_{\rm AB}^2 M^2}{r_a^2}\left(\nu [f_{11} -  f_{12} - f_{13} +  f_{14}]+\frac{f_{13}m_1}{M}+\frac{f_{12}m_2}{M}\right)-\frac{G_{\rm AB}M}{r_a}\nu \left(\frac{1}{8}+3 f_{10}+ 3 f_7+3 \frac{m_1}{M}f_9+3 \frac{m_2}{M}f_8\right)\dot{r}_a^2\right.\right.
		\nonumber\\
		&\left.\left.-\frac{G_{\rm AB}M}{r_a}\left(-\frac{7}{8}\nu-\frac{1}{2}\nu \gamma_{\rm AB}+f_1 \nu-f_3 \nu -f_4\nu +f_6\nu -\frac{m_1}{M}f_6-\frac{m_2}{M}f_1+\frac{m_1}{M}\nu (f_4+f_5+f_6)+\frac{m_2}{M}\nu (f_1+ f_2+ f_3)\right)v_a^2\right]\right.
		\nonumber\\
		&\left.-\textbf{v}_a\frac{G_{\rm AB}M}{r_a}\nu\left[\frac{7}{4}+\gamma_{\rm AB}+2 (f_1+f_6)+2\frac{m_1}{M}(f_5-f_3)+2\frac{m_2}{M}(f_2-f_4)\right] r_a \dot{r}_a\right\}+\mathcal{O}\left(\frac{1}{c^5}\right)~, \nonumber\\
		\textbf{v}_h &= \frac{d \textbf{r}_h}{dt}= \textbf{v}_a + \frac{1}{c^4}\left\{\frac{G_{\rm AB}^2M^2}{r_a^2}\left[\textbf{v}_a\left(\frac{7}{4}\nu+\nu(2f_1+2f_6+f_{11}-f_{12}-f_{13}+f_{14})+\frac{m_2}{M}(f_{12}+2\nu f_2 -2 \nu f_4)\right.\right.\right.
		\nonumber\\
		&\left.\left.\left.+\frac{m_1}{M}(f_{13}+2f_5\nu-2 f_3 \nu)\right)+\frac{\textbf{r}_a}{r_a}\dot{r}_a \left(-\frac{\nu}{4}-2\nu(2f_1+3f_{10}-f_{11}+f_{12}+f_{13}-f_{14}-f_3-f_4+2f_6+3f_7)\right.\right.\right.
		\nonumber\\
		&\left.\left.\left.-\frac{m_1}{M}(2 [f_{13}+f_6]-\nu[-2 f_3+2 f_4+4 f_5+2 f_6+6 f_9])-\frac{m_2}{M}(2 [f_{12}+f_1]-\nu[-2 f_4+2 f_1+4 f_2+2 f_3+6 f_8])\right)\right]\right.
		\nonumber\\
		&\left.+\frac{G_{\rm AB}M}{r_a}\left[\textbf{v}_a \left(-\frac{7}{8}\nu+\nu[-3 f_1 + f_3 +f_4-3 f_6]+\frac{m_1}{M}[f_6 + \nu (2 f_3-f_4-3 f_5 -f_6)] + \frac{m_2}{M}[f_1+\nu(2f_4-f_3-3 f_2 -f_1)]\right)v_a^2\right.\right.
		\nonumber\\
		&\left.\left.+\textbf{v}_a \nu \left(\frac{13}{8}+2f_1-3f_{10}+2f_6-3f_7+\frac{m_1}{M}[-2f_3+2f_5-3f_9]+\frac{m_2}{M}[2f_2-2f_4-3f_8]\right)\dot{r}_a^2+\frac{\textbf{r}_a}{r_a}\nu\left(\frac{3}{8}+9 f_{10}+9 f_7 + 9\frac{m_1}{M}f_9\right.\right.\right.
		\nonumber\\
		&\left.\left.\left. + 9\frac{m_2}{M}f_8\right)\dot{r}_a^3 +\frac{\textbf{r}_a}{r_a}\left(-\frac{9}{8}\nu+\nu[f_1-6f_{10}-f_3-f_4+f_6-6f_7]-\frac{m_1}{M}[f_6-\nu(f_4+f_5+f_6-6f_9)]\right.\right.\right.
		\nonumber\\
		&\left.\left.\left.-\frac{m_2}{M}[f_1-\nu(f_1+f_2+f_3-6f_8)]\right)v_a^2\dot{r}_a\right]\right\}+\mathcal{O}\left(\frac{1}{c^5}\right)~,\nonumber\\
		\frac{\textbf{p}_a}{\mu}& = \textbf{v}_a +\frac{1}{c^2}\left[\nu\frac{G_{\rm AB}M}{r_a} \dot{r}_a~\textbf{n}_a + v_a^2\frac{(1-3\nu)}{2}\textbf{v}_a+(3+ \nu+ 2\gamma_{\rm AB})\frac{G_{\rm AB}M}{r_a}\textbf{v}_a\right]+\frac{1}{c^4}\left\{\frac{3}{8}\Bigg(1-7\nu+13\nu^2\Bigg) v_a^4 \textbf{v}_a\right.
		\nonumber\\
		&\left.+\frac{G_{\rm AB}M}{2r_a}\Bigg[\Big(7-10\nu-9\nu^2-8(f_1+f_6)\nu+4(1-2\nu)\gamma_{\rm AB}+8\frac{m_1}{M}\left(f_6+[f_3-f_5-f_6]\nu\right)+8\frac{m_2}{M}\left(f_1+[f_4-f_1-f_2]\nu\right)\Big)v_a^2 \right.
		\nonumber\\
		&\left.+\Big([1-5\nu+4f_1-12f_{10}+4f_6-12f_7]\nu+4\frac{m_1}{M}\left[3f_{10}(1-\nu)-f_6+(f_5+f_6-f_3-3f_9)\nu\right]+4\frac{m_2}{M}\left[3f_{7}(1-\nu)-f_1\right.\right.
		\nonumber\\
		&\left.\left.+(f_1+f_2-f_4-3f_8)\nu\right]\Big)\dot{r}_a^2\Bigg]\textbf{v}_a+\frac{G_{\rm AB}M}{2r_a}\dot{r}_a\Bigg[\Big([1+4f_1+4f_6-12f_7-12f_{10}-5\nu]\nu+4\frac{m_1}{M}\left[3f_{10}(1-\nu)-f_6(1-\nu)\right.\right.
		\nonumber\\
		&\left.\left.+(f_5-f_3-3f_9)\nu\right]+4\frac{m_2}{M}\left[3f_{7}(1-\nu)-f_1(1-\nu)+(f_2-f_4-3f_8)\nu\right]\Big)v_a^2+\Big(24f_{10}\nu+24f_7\nu+3\nu^2\right.
		\nonumber\\
		&\left.+24\frac{m_1}{M}\left[-f_{10}+(f_9+f_{10})\nu\right]+24\frac{m_2}{M}\left[-f_7+(f_7+f_8)\nu\right]\Big)\dot{r}_a^2\Bigg]\textbf{n}_a+\frac{G_{\rm AB}^2M^2}{r_a^2}\Bigg[\frac{7}{2}-\frac{17}{2}\nu+\nu^2+\frac{11}{4}\gamma_{\rm AB}^2-5\gamma_{\rm AB}(-1+\nu)\right.
		\nonumber\\
		&\left.+\langle\beta\rangle(1-\nu)+2\nu\beta_+-\langle\delta\rangle+2(f_1-f_{12}-f_{13}+f_6)\nu+2\frac{m_1}{M}\left(f_{13}+f_6[-1+\nu]+[f_{11}+f_{14}-f_3+f_5]\nu\right)\right.
		\nonumber\\
		&\left.+2\frac{m_2}{M}\left(f_{12}+f_1[-1+\nu]+[f_{11}+f_{14}-f_4+f_2]\nu\right)\Bigg]\textbf{v}_a+\frac{G_{\rm AB}^2M^2}{4r_a^2}\dot{r}_a\Bigg[8\left(7+5\gamma_{\rm AB}-\beta_- - 3\beta_+ +2[f_1+f_{12}+f_{13}+f_6]\right.\right.
		\nonumber\\
		&\left.\left.+3[f_{10}+f_7]\right)\nu+12\nu^2+4\langle\delta\rangle+(2+\gamma_{\rm AB})^2+8\frac{m_1}{M}\left(-3f_{10}-2[f_{13}+f_6]+[3_{10}-2(f_{11}+f_{14}+f_3-f_5-f_6)+3 f_9]\nu\right)\right.
		\nonumber\\
		&\left.+8\frac{m_2}{M}\left(-3f_{7}-2[f_{12}+f_1]+[3_{7}-2(f_{11}+f_{14}+f_4-f_2-f_1)+3 f_8]\nu\right)\Bigg]\textbf{n}_a\right\}+ \mathcal{O}\left(\frac{1}{c^6}\right)~,\nonumber\\
		\textbf{v}_a & = \tilde{\textbf{p}}_a +\frac{1}{c^2}\left(-\nu\frac{G_{\rm AB}M}{r_a} \tilde{p}_{r,a}~\textbf{n}_a - \tilde{p}_a^2\frac{(1-3\nu)}{2}\tilde{\textbf{p}}_a-(3+ \nu+ 2\gamma_{\rm AB})\frac{G_{\rm AB}M}{r_a}\tilde{\textbf{p}}_a\right) + \frac{1}{c^4}\left\{\frac{3}{8}\Big(1-5\nu+5\nu^2\Big) \tilde{p}_a^4 \tilde{\textbf{p}}_a \right.
		\nonumber\\
		&\left. +\frac{G_{\rm AB}M}{r_a}\Bigg[\frac{3}{8}\Bigg(\nu-4\nu^2-8f_{10}\nu-8f_7\nu-\frac{m_1}{M}[8f_{10}(-1+\nu)+8f_9\nu]-\frac{m_2}{M}[8f_{7}(-1+\nu)+8f_8\nu]\Bigg)\tilde{\textbf{p}}_{r, a}^2-\frac{1}{8}\Bigg(5\nu+4\nu^2\right.
		\nonumber\\
		&\left.+8f_1\nu+8f_6\nu+\frac{m_1}{M}[8f_6(-1+\nu)+(-8f_3+8f_5)\nu]+\frac{m_2}{M}[8f_1(-1+\nu)+(-8f_4+8f_2)\nu]\Bigg)\tilde{p}_a^2\Bigg]\tilde{\textbf{p}}_{r, a}\textbf{n}_a\right.
		\nonumber\\
		&\left.+\frac{G_{\rm AB}M}{r_a}\Bigg[\frac{1}{8}\Bigg(17\nu-4\nu^2+24[f_7+f_{10}]\nu+24\frac{m_1}{M}[f_9\nu+(-1+\nu)f_{10}]+24\frac{m_2}{M}[f_8\nu+(-1+\nu)f_{7}]\Bigg)\tilde{p}_{r,a}^2+\frac{1}{8}\Bigg(20-95\nu\right.
		\nonumber\\
		&\left.-12\nu^2+\gamma_{\rm AB}[16-64\nu]+8f_1\nu+8f_6\nu+8\frac{m_1}{M}[f_5\nu-f_3\nu+(-1+\nu)f_6]+8\frac{m_2}{M}[f_2\nu-f_4\nu+(-1+\nu)f_1]\Bigg)\tilde{p}_a^2\Bigg]\tilde{\textbf{p}}_a\right.
		\nonumber\\
		&\left.-\frac{G_{\rm AB}^2 M^2}{4r_a^2}\Bigg[\Bigg(4+31\nu+4\gamma_{\rm AB}[1+6\nu]+\gamma_{\rm AB}^2-16\langle\beta\rangle \nu-8[\beta_+ + X_{\rm AB} \beta_-]\nu + 4 \langle\delta\rangle+8[f_{12}+f_{13}-f_{11}-f_{14}]\nu-8\frac{m_1}{M}f_{13}\right.
		\nonumber\\
		&\left.-8\frac{m_2}{M}f_{12}\Bigg)\tilde{p}_{r,a}\textbf{n}_a-\Bigg(22+65\nu+5\gamma_{\rm AB}^2+4\gamma_{\rm AB}[7+9\nu]-4\langle\beta\rangle[1+\nu]-8\nu X_{\rm AB}\beta_-+4\langle\delta\rangle -4[f_{11}-f_{12}-f_{13}+f_{14}]\nu\right.
		\nonumber\\
		&\left.-4\frac{m_1}{M}f_{13}-4\frac{m_2}{M}f_{12}\Bigg)\tilde{\textbf{p}}_a\Bigg]\right\}+ \mathcal{O}\left(\frac{1}{c^6}\right)~,\nonumber\\
		\textbf{n}_h &= \textbf{n}_a+\frac{1}{c^4}\left\{\nu\frac{G_{\rm AB} M}{r_a}\left[\frac{7}{4}+ \gamma_{\rm AB}+\frac{f_1}{2}+\frac{f_6}{2}+\frac{m_1}{M}\left(-2 f_3+2f_5\right)+\frac{m_2}{M}\left(2 f_2-2f_4\right)\right] \left( \textbf{n}_a \dot{r}_a^2- \textbf{v}_a \dot{r}_a\right)\right\} + \mathcal{O}\left(\frac{1}{c^5}\right)~.
	\end{align}
\end{widetext}	
Therefore, the explicit transformation of the fundamental scalar quantities from the harmonic coordinate system to the ADM coordinate system up to 2PN order is
\begin{widetext}
	\begin{align}
		v_h^2&= \tilde{p}_a^2+\frac{1}{c^2}\left(-2\nu\frac{G_{\rm AB}M}{r_a} \tilde{p}_{r, a}^2 -(1-3\nu) \tilde{p}_a^4-2(3+ \nu+ 2\gamma_{\rm AB})\frac{G_{\rm AB}M}{r_a}\tilde{p}_a^2\right)+\frac{1}{c^4}\Bigg\{\frac{1}{4}\left(4-21\nu+24\nu^2\right)\tilde{p}_a^6
		\nonumber\\
		&+\frac{G_{\rm AB}M}{r_a}\Bigg[\Bigg(8-6\nu^2+\gamma_{\rm AB}[6-22\nu]-\frac{127}{4}\nu+2[f_1+f_6]+2\frac{m_1}{M}[f_6(-1+\nu)-f_3+f_5]+2\frac{m_2}{M}[f_1(-1+\nu)-f_4+f_2]\Bigg)\tilde{p}_a^4
		\nonumber\\
		&+\Bigg(4\nu-5\nu^2+6f_{10}\nu-2f_6\nu-2[f_1+f_3+f_4-f_7]\nu+2\frac{m_1}{M}[f_6+3f_{10}(-1+\nu)-f_6\nu+(2f_3+f_4-f_5+3f_9)\nu]
		\nonumber\\
		&+2\frac{m_2}{M}[f_1+3f_{7}(-1+\nu)-f_1\nu+(2f_4+f_3-f_2+3f_8)\nu]\Bigg)\tilde{p}_{r,a}^2\tilde{p}_a^2-\frac{3}{4}\Bigg(-\nu+4\nu^2+8[f_{10}+f_{7}]\nu+8\frac{m_1}{M}[f_9\nu+f_{10}(\nu-1)]
		\nonumber\\
		&+8\frac{m_2}{M}[f_8\nu+f_{7}(\nu-1)]\Bigg)\tilde{p}_{r,a}^4\Bigg]+\frac{G_{\rm AB}^2 M^2}{r_a^2}\Bigg[\Bigg(-2-\frac{\gamma_{\rm AB}^2}{2}-2\langle\delta\rangle-2\gamma_{\rm AB}(1+4\nu)+\nu[-\frac{19}{2}+3\nu+4\beta_+ -4 \beta_- X_{\rm AB}+8 \langle\beta \rangle
		\nonumber\\
		&+4f_{11}-4f_{12}-4f_{13}+4f_{14}]\nu+4\frac{m_1}{M}f_{13}+4\frac{m_2}{M}f_{12}\Bigg)\tilde{p}_{r,a}^2+\Bigg(20+\frac{13}{2}\gamma_{\rm AB}^2+2\langle\delta\rangle+\frac{77}{2}\nu+4\nu \beta_- X_{\rm AB}-2\nu\beta_++\nu^2
		\nonumber\\
		&-2\langle\beta\rangle[1+\nu]+\gamma_{\rm AB}[26+22\nu]-2f_{11}\nu+2f_{12}\nu+2f_{13}\nu-2f_{14}\nu-2\frac{m_1}{M}f_{13}-2\frac{m_2}{M}f_{12}\Bigg)\tilde{p}_a^2\Bigg]\Bigg\}+\mathcal{O}\left(\frac{1}{c^6}\right)~,\nonumber\\
		\dot{r}_h^2&=\tilde{p}_{r,a}^2+\frac{1}{c^2}\left(-2 \nu \frac{G_{\rm AB}M}{r_a}\tilde{p}_{r, a}^2-\tilde{p}_a^2(1-3\nu)\tilde{p}_{r, a}^2- 2(3+\nu+2\gamma_{\rm AB})\frac{G_{\rm AB}M}{r_a}\tilde{p}_{r, a}^2\right)+\frac{1}{c^4}\Bigg\{\frac{1}{4}\left(4-21\nu+24\nu^2\right)\tilde{p}_{r, a}^2 \tilde{p}_{a}^4
		\nonumber\\
		&-\frac{G_{\rm AB}M}{2r_a}\Bigg[-16+71\nu+20\nu^2+4\nu\gamma_{\rm AB}+4\gamma_{\rm AB}(-3+11\nu)+8(f_1+f_6)\nu+8\frac{m_1}{M}\Big(-f_6+[-f_3+f_5+f_6]\nu\Big)
		\nonumber\\
		&+8\frac{m_2}{M}\Big(-f_1+[-f_4+f_1+f_2]\nu\Big)\Bigg]\tilde{p}_{r,a}^2\tilde{p}_a^2+\frac{G_{\rm AB}M}{2r_a}\Bigg[17\nu+4\nu\gamma_{\rm AB}-8\nu^2+8[f_1+f_6]\nu+8\frac{m_1}{M}\Bigg(-f_6+[-f_3+f_4+f_5]\nu\Bigg)
		\nonumber\\
		&+8\frac{m_2}{M}\Bigg(-f_1+[-f_4+f_2+f_1]\nu\Bigg)\Bigg]\tilde{p}_{r,a}^4+\frac{G_{\rm AB}^2M^2}{r_a^2}\Bigg[\Bigg(29+14\gamma_{\rm AB}+6\langle\beta\rangle+2\beta_+ + 2f_{11}-2f_{12}-2f_{13}+2f_{14}\Bigg)\nu+4\nu^2-2\langle\beta\rangle
		\nonumber\\
		&+6(1+\gamma_{\rm AB})(3+\gamma_{\rm AB})+2\frac{m_1}{M}f_{13}+2\frac{m_2}{M}f_{12}\Bigg]\Bigg\}+\mathcal{O}\left(\frac{1}{c^4}\right)~,\nonumber\\
		\frac{1}{r_h}& = \frac{1}{r_a}\Bigg\{1+\frac{1}{c^4}\Bigg[\frac{G_{\rm AB}M}{8r_a}\Big([-7-4\gamma_{\rm AB}+8f_1-8f_3-8f_4+8f_6]\nu+8\frac{m_1}{M}[-f_6+(f_4+f_5+f_6)\nu]+8\frac{m_2}{M}[-f_1
		\nonumber\\
		&+(f_1+f_2+f_3)\nu]\Big)\tilde{p}_a^2+\frac{G_{\rm AB}M}{8r_a}\Big([15+8\gamma_{\rm AB}+16f_1+24f_{10}+16f_6+24f_7]\nu+8\frac{m_1}{M}[(2f_6+3f_{10})(\nu-1)
		\nonumber\\
		&+(2f_5-2f_3+3f_9)\nu]+8\frac{m_2}{M}[(2f_1+3f_{7})(\nu-1)+(2f_2-2f_4+3f_8)\nu]\Big)\tilde{p}_{r,a}^2\Bigg]-\frac{G_{\rm AB}^2M^2}{r_a^2}\Bigg[\Big(f_{11}-f_{12}-f_{13}+f_{14}\Big)\nu
		\nonumber\\
		&+\frac{m_1}{M}f_{13}+\frac{m_2}{M}f_{12}\Bigg]\Bigg\}+\mathcal{O}\left(\frac{1}{c^6}\right)~.
	\end{align}
\end{widetext}
Here, $f_i$'s are the parameters of the total derivative function $f$ entering at 2PN order \cite{Julie:2017pkb} and
\begin{align*}
&\langle\bar{\beta}\rangle\equiv -X_{AB}\beta_-+\beta_+,\\
&\langle{\delta}\rangle\equiv X_{AB}\delta_-+\delta_+~,\\
\end{align*}
with $X_{\rm AB}\equiv (m_A-m_B)/M \ $ and $``{\pm}''$ subscript denoting the symmetric and anti-symmetric parts of the ST parameters, e.g. \hbox{$x_{\pm}\equiv(x_A\pm x_B)/2$} \cite{Jain:2022nxs}. From the above transformations, we verify that the harmonic to ADM coordinate transformation of scalar-tensor theory reduces to the coordinate transformation of GR in the GR limit \cite{Bini:2012ji}.
	
	Until now, we only computed the transformation involving the conservative orders up to 2PN order. However, there still remains the possible transformation (gauge choice) in the radiation reaction terms which enter at 1.5PN order due to the scalar dipolar radiation in the scalar-tensor theory. The lowest order, i.e. 1.5PN order dissipative relative acceleration term in harmonic coordinates is \cite{Bernard:2022noq} 
	\begin{align}
	\label{eqn:rr-harmonic}
	\textbf{a}^{\mathrm{1.5PN}}_h = \frac{4\zeta\mathcal{S}_-^2}{3c^3r_h}\frac{G_{\rm AB}^2 M^2 \nu}{r_h^2}\left[3 \dot{r}_h \textbf{n}_h-\textbf{v}_h\right]~.
	\end{align}
	Now, to investigate the alternative gauge choices for radiation reaction at the leading order, we first write the general form of the radiation reaction term for the binary system without spins and tidal effects using the fact that: (i) the term must be a correction to the Newtonian acceleration implying that it should be proportional to $(GM)/r^2$, (ii) should vanish in the test-mass limit, i.e. should be proportional to symmetric mass-ratio $\nu$, (iii) must be non-linear in Newton's constant as it is related to the emission of gravitational radiation [i.e., should contain an extra factor of $(GM)/r$], and finally, (iv) must be odd in velocities, i.e. dissipative. As we are working in scalar-tensor theories, Newton's constant is replaced by the coupling constant $G_{\rm AB}$ of the theory. Therefore, the general radiation reaction term at the leading order reads
	\begin{align}
	\label{eqn:rr-general}
	\textbf{a}_{\mathrm{1.5PN}} = \frac{G_{\rm AB}^2 M^2}{r^3}\nu \left[A_{3/2}\, \dot{r}\, \textbf{n} + B_{3/2}\, \textbf{v}\right]~,
	\end{align}	
	where $v$, $n$ denote the relative velocity and separation vector. The values for the parameters $A_{3/2}$ and $B_{3/2}$ in harmonic coordinates can be read from Eq.~\eqref{eqn:rr-harmonic}. The goal of our investigation would now be to determine these two arbitrary parameters in ADM coordinates, for which it will be sufficient to investigate for gauge freedom in defining radiation-reaction terms.
	
	For this, we take the post-Newtonian expressions for the energy and angular momentum, $\tilde{E}\equiv E/\mu$ and $\tilde{J}\equiv J/\mu$, and calculate $d\tilde{E}/dt$ and $d\tilde{J}/dt$ using 1PN equations of motion supplemented with 1.5PN radiation reaction of Eq.~\eqref{eqn:rr-general}. The angular momentum and energy are conserved up to 1PN order but the presence of dissipative effects at 1.5PN order leads to non-vanishing values of  $d\tilde{E}/dt$ and $d\tilde{J}/dt$, and hence we have the freedom to add arbitrary terms to $\tilde{E}$ and $\tilde{J}$ at 1.5PN order (see Ref.~\cite{Iyer:1995rn} for similar analysis). However, the term should be such that it is proportional to $\nu$ (vanishes in the test-mass limit), $(GM)/r$ (non-linear effects for gravitational waves emission), and odd in $\dot{r}$ (dissipative effects). Therefore, the possible terms for $\tilde{E}$ and $\tilde{J}$ reduce to
	\begin{align}
	\tilde{E}^*\equiv\tilde{E}_N +\nu~\dot{r}\left(\frac{G_{\rm AB}M}{r}\right)^2  \alpha_{3/2}~,\nonumber\\
	\tilde{J}^*\equiv\tilde{J}_N + \nu~\dot{r}\tilde{J}_N\left(\frac{G_{\rm AB}M}{r}\right) \beta_{3/2}~,
	\end{align}
	where we use `*' to distinguish the arbitrary terms entering at 1.5PN order from the conserved quantities up to 1PN order, $\tilde{E}_N $ and $\tilde{J}_N=\textbf{x}\times\textbf{v}$ are the Newtonian order contribution to the energy and angular momentum, $\alpha_{3/2}$ and $\beta_{3/2}$ are the arbitrary parameters entering at 1.5PN order. Therefore, the energy and angular momentum orbital loss reads
	\begin{align}
\frac{d\tilde{E}^*}{dt}&=\textbf{v}\cdot\textbf{a}_{\rm 1.5PN}+ \nonumber \\
&\quad+\nu\, \alpha_{3/2}\left[\ddot{r}\left(\frac{G_{\rm AB}M}{r}\right)^2-2\frac{\dot{r}^2}{r}\left(\frac{G_{\rm AB}M}{r}\right)^2\right]\,,\nonumber\\
\frac{d\tilde{J}^*}{dt}&=\textbf{x}\times\textbf{a}_{\rm 1.5PN}+\nu\,\beta_{3/2}\tilde{J}_N\left[\frac{G_{\rm AB}M}{r}\ddot{r}-\frac{G_{\rm AB}M}{r}\frac{\dot{r}^2}{r}\right]\,.
	\end{align}
	Now, in order to find the expression of the derivatives, we use the Newtonian order equations of motion, i.e.
	\begin{align}
	\ddot{r} = -\frac{G_{\rm AB}M}{r^2}+\frac{v^2}{r}-\frac{\dot{r}^2}{r}~,
	\end{align}
	and acceleration at 1.5PN order from Eq.~\eqref{eqn:rr-general}. Hence, the explicit expression of the energy and angular momentum loss reads
	\begin{align}
		\label{eqn:generalfluxEJ}
	\frac{d\tilde{E}^*}{dt}=&\left(\frac{G_{\rm AB}M}{r}\right)^2\frac{\nu}{r}\left[A_{3/2}\dot{r}^2+B_{3/2}v^2\right.\nonumber\\
	&\left.+\alpha_{3/2}\left(v^2-\frac{G_{AB}M}{r}-3\dot{r}^2\right)\right]~,\nonumber \\
		\frac{d\tilde{J}^*}{dt}=&\frac{G_{\rm AB}M}{r^2}\frac{\nu\tilde{J}_N}{r}\beta_{3/2}\left(v^2-\frac{G_{AB}M}{r}-2\dot{r}^2\right)\nonumber\\
		&+\left(\frac{G_{\rm AB}M}{r}\right)^2\frac{\nu\tilde{J}_N}{r}B_{3/2}~.
	\end{align}
	We now use that the energy and angular momentum balance equations, i.e. the angular momentum and the energy orbital loss corresponds to the far-zone fluxes
	\begin{align}
	\frac{d\tilde{E}^*}{dt} = -\left(\frac{d\tilde{E}}{dt}\right)_{\rm far-zone}, \nonumber\\
	\frac{d\tilde{J}^*}{dt} = -\left(\frac{d\tilde{J}}{dt}\right)_{\rm far-zone},
	\end{align}
	where the far-zone fluxes are
	\begin{align}
	\label{eqn:far-zonefluxEJ}
	\left(\frac{d\tilde{E}}{dt}\right)_{\rm far-zone}&=\frac{4}{3}\,\zeta \,\mathcal{S}_-^2\,\frac{\nu}{r}\left(\frac{G_{\rm AB}M}{r}\right)^3~,\nonumber\\
	\left(\frac{d\tilde{J}}{dt}\right)_{\rm far-zone}&=\frac{4}{3}\,\zeta \,\mathcal{S}_-^2\,\frac{\nu}{r}\left(\frac{G_{\rm AB}M}{r}\right)^2\tilde{J}_N~.
	\end{align}
	Comparing term-by-term Eqs.~\eqref{eqn:generalfluxEJ} and~\eqref{eqn:far-zonefluxEJ}, we see that all the four arbitrary parameters: $A_{3/2}$, $B_{3.2}$, $\alpha_{3/2}$ and $\beta_{3/2}$ are uniquely determined, namely
	\begin{align}
	\label{eq:pars}
	A_{3/2}&=4\zeta \mathcal{S}_-^2~,\hspace{0.3cm}\alpha_{3/2}=\frac{4}{3}\zeta \mathcal{S}_-^2~,\nonumber\\
	B_{3/2}&=-\frac{4}{3}\zeta \mathcal{S}_-^2~,\hspace{0.3cm}\beta_{3/2}=0~.
	\end{align}
The far-zone energy and angular momentum fluxes expressed in terms of the radiative multipole moments at infinity are gauge invariant, i.e. it is independent of the coordinate system that preserves the asymptotic flatness. Thus, the unique determination of the arbitrary coefficients implies that there is no residual gauge freedom at 1.5PN order in scalar-tensor theories in contrast to the residual gauge freedom at 2.5PN order in GR (see, Refs.~\cite{Iyer:1993xi, Iyer:1995rn}). Furthermore, this implies that there is no coordinate transformation between ADM and harmonic coordinates at 1.5PN order, and hence no coordinate gauge freedom in the radiation reaction force.
	
	\subsubsection{ADM to EOB}
	\label{sec:admtoeob}
	In this section, we derive the transformation of the fundamental scalars from ADM coordinates to EOB coordinates up to 2PN order.
	The ADM to EOB canonical transformation up to 3PN order is derived in Refs.~\cite{Julie:2017pkb,Jain:2022nxs}. For the purpose of this work, we consider transformation only up to 2PN order. 

	The explicit transformation of the fundamental scalar quantities in ADM coordinate system to the EOB coordinate system reads
	\begin{widetext}
	\begin{align}
		\frac{1}{r_a} &= \frac{1}{r_e}+ \frac{1}{c^2}\Bigg\{\frac{1}{r_e}\left[-\frac{\nu}{2}\tilde{p}_e^2+\frac{G_{\rm AB}M}{r_e}\left(1+\gamma_{\rm AB}+\frac{\nu}{2}\right)\right]-\frac{\nu}{ r_e} \tilde{p}_{r, e}^2\Bigg\}+\frac{1}{r_e c^4}\Bigg\{\Big[\nu+3\nu^2\Big]\frac{\tilde{p}_e^4}{8}+\frac{3\nu^2}{2}\tilde{p}_{r,e}^4+\frac{\nu}{2}\tilde{p}_e^2\tilde{p}_{r,e}^2
		\nonumber\\
		&+\frac{G_{\rm AB}M}{r_e}\Bigg[\Big(-\frac{7}{8}\nu^2-f_6\nu-f_1\nu+\frac{m_1}{M}[f_6+(f_3-f_5-f_6\nu)]+\frac{m_2}{M}[f_1+(f_4-f_1-f_2\nu)]\Big)\tilde{p}_e^2-\Big(-3\nu+\frac{5}{8}\nu^2-3\gamma_{\rm AB}\nu
		\nonumber\\
		&+2[f_1+f_6]\nu+3[f_{10}+f_7]\nu+\frac{m_1}{M}[(3f_{10}+2f_6)(\nu-1)+(3f_9+2f_5-2f_3)\nu]+\frac{m_2}{M}[(3f_{7}+2f_1)(\nu-1)
		\nonumber\\
		&+(3f_8+2f_2-2f_4)\nu]\Big)\tilde{p}_{r, e}^2\Bigg]+\frac{G_{\rm AB}^2M^2}{r_e^2}\Bigg[1-\frac{15}{4}\nu+\frac{\nu^2}{2}+\gamma_{\rm AB}\left(\frac{5}{2}-2\nu\right)+\frac{11}{8}\gamma_{\rm AB}^2+\langle\beta\rangle+\frac{\nu}{2}\beta_+-\frac{1}{2}\langle\delta\rangle
		\nonumber\\
		&+(f_{11}-f_{12}-f_{13}+f_{14})\nu+\frac{m_1}{M}f_{13}+\frac{m_2}{M}f_{12}\Bigg]\Bigg\}+\mathcal{O}\left(\frac{1}{c^6}\right),\nonumber\\
		p_{r, a}^2 &= p_{r, e}^2 - \frac{1}{c^2}\left[\nu p_{r, e}^2\left(2\tilde{p}_{r, e}^2-\tilde{p}_e^2\right)\right]+\frac{p_{r,e}^2}{c^4}\Bigg\{4\nu^2 \tilde{p}_{r,e}^4-\frac{3\nu}{4}\tilde{p}_e^4+\frac{G_{\rm AB}^2M^2}{r_e^2}\Bigg[-\frac{19}{2}\nu+\frac{1}{2}\nu^2+\gamma_{\rm AB}(1-6\nu)+\frac{3}{4}\gamma_{\rm AB}^2+2\langle\beta\rangle+\nu\beta_+ 
		\nonumber\\
		&-\langle\delta\rangle+2(f_{11}-f_{12}-f_{13}+f_{14})\nu-2\frac{m_1}{M}f_{13}-2\frac{m_2}{M}f_{12}\Bigg]+\nu\Big(1-3\nu) \tilde{p}_e^2\tilde{p}_{r,e}^2 +\frac{G_{\rm AB}M}{r_e}(\tilde{p}_{r,e}^2-\tilde{p}_e^2)\Bigg[-4(-1+f_1+f_6-\gamma_{\rm AB})\nu
		\nonumber\\
		& -\frac{3}{2}\nu^2+4\frac{m_1}{M}\Big(f_6-[f_5+f_6-f_3]\nu\Big)+4\frac{m_2}{M}\Big(f_1-[f_1+f_2-f_4]\nu\Big)\Bigg]+\mathcal{O}\left(\frac{1}{c^6}\right),\nonumber\\
		p_a^2 &= p_e^2 + \frac{1}{c^2}\Bigg[\frac{G_{\rm AB}M}{r_e}(p_e^2-p_{r, e}^2)\left(2+\nu+2\gamma_{\rm AB}\right)-\nu p_e^2 \tilde{p}_e^2\Bigg]+\frac{1}{c^4}\Bigg\{\Big[\frac{1}{4}\nu+\nu^2\Big]p_e^2\tilde{p}_e^4+\frac{G_{\rm AB}M}{4r_e}(p_{r,e}^2-p_e^2)\Bigg[\tilde{p}_e^2\Big(4\nu+9\nu^2+4\gamma_{\rm AB}
		\nonumber\\
		&+8[f_1+f_6]\nu+8\frac{m_1}{M}[f_6(\nu-1)-f_3\nu+f_5\nu]+8\frac{m_2}{M}[f_1(\nu-1)-f_4\nu+f_2\nu]\Big)+\tilde{p}_{r,e}^2\Big(3\nu^2+24[f_{10}+f_7]\nu
		\nonumber\\
		&+24\frac{m_1}{M}[f_9\nu+f_{10}(\nu-1)]+24\frac{m_2}{M}[f_8\nu+f_{7}(\nu-1)]\Big)\Bigg]+\frac{G_{\rm AB}^2M^2}{4r_e^2}p_{e}^2\Bigg[12-26\nu+5\nu^2+4\gamma_{\rm AB}(7-3\nu)+15\gamma_{\rm AB}^2+8\langle\beta\rangle
		\nonumber\\
		&+4\nu\beta_+-4\langle\delta\rangle+8\nu(f_{11}-f_{12}-f_{13}+f_{14})+8\frac{m_1}{M}f_{13}+8\frac{m_2}{M}f_{12}\Bigg] +\frac{G_{\rm AB}^2M^2}{4r_e^2}p_{r,e}^2\Bigg[-12+64\nu-7\nu^2-32\gamma_{\rm AB}+36\gamma_{\rm AB}\nu
		\nonumber\\
		&-18\gamma_{\rm AB}^2-16\langle\beta\rangle-8\nu\beta_++8\langle\delta\rangle-16\nu(f_{11}+f_{14}-f_{12}-f_{13})-16\frac{m_1}{M}f_{13}-16\frac{m_2}{M}f_{12}\Bigg]\Bigg\}+\mathcal{O}\left(\frac{1}{c^6}\right).
	\end{align}
	\end{widetext}
	\subsubsection{Harmonic to EOB}
As we want to transform the harmonic coordinate angular momentum flux and energy flux presented in Refs.~\cite{Jain:2024lie,Bernard:2022noq,Lange:2017wki,Lange:2018pyp}, we combine the two transformation given in sections \ref{sec:htoadm}-\ref{sec:admtoeob} to obtain the transformation from harmonic coordinates to EOB coordinates. The explicit expression of the transformation up to 2PN order reads 
	\begin{align}
		\dot{r}_h^2&=\tilde{p}_{r, e}^2+\frac{1}{c^2}\Bigg[-2\nu \tilde{p}_{r, e}^4+(4\nu-1)\tilde{p}_e^2\tilde{p}_{r, e}^2\nonumber\\
		&\quad-2(3+2\nu+2\gamma_{\rm AB})\frac{G_{\rm AB}M}{r_e}\tilde{p}_{r, e}^2\Bigg]+\frac{1}{c^4}\Bigg[4\nu^2 \tilde{p}_{r, e}^6\nonumber\\
		&+(3\nu-9\nu^2)\tilde{p}_{r, e}^4 \tilde{p}_{e}^2+(1-6\nu+6\nu^2)\tilde{p}_{r, e}^2\tilde{p}_{e}^4\nonumber\\
		&+\Bigg(2+\frac{51}{2}\nu+\frac{7}{2}\nu^2+2\gamma_{\rm AB}+12\gamma_{\rm AB}\nu\Bigg)\frac{G_{\rm AB} M}{r_e} \tilde{p}_{r, e}^4 \nonumber\\
		&+\Bigg(6-\frac{75}{2}\nu-\frac{15}{2}\nu^2+4\gamma_{\rm AB}-24\gamma_{\rm AB}\nu\Bigg)\frac{G_{\rm AB} M}{r_e} \tilde{p}_{r, e}^2\tilde{p}_{e}^2\nonumber\\
		&+\Bigg(12+\frac{63}{2}\nu+\frac{3}{2}\nu^2+13\gamma_{\rm AB}+\frac{5}{4}\gamma_{\rm AB}^2+14\gamma_{\rm AB}\nu\nonumber\\
		&+3\langle\beta\rangle(-1+\nu)+\beta_+\nu+\langle\delta\rangle\Bigg)\frac{G_{\rm AB}^2 M^2}{r_e^2} \tilde{p}_{r, e}^2\Bigg]+\mathcal{O}\left(\frac{1}{c^5}\right),\nonumber\\
		v_h^2&= \tilde{p}_e^2+\frac{1}{c^2}\Bigg[\left(-1+2\nu\right)\tilde{p}_e^4
		 +\frac{G_{\rm AB}M}{r_e}\Bigg(\tilde{p}_{r, e}^2(-2-3\nu
		 \nonumber\\
		&-2\gamma_{\rm AB})
		 -\tilde{p}_e^2(4+\nu+2\gamma_{\rm AB})\Bigg)\Bigg] +\frac{1}{c^4}\Bigg\{(1-3\nu+\nu^2)\tilde{p}_e^6\nonumber\\
		&+\frac{G_{\rm AB}M}{r_e}\Bigg[\Big(4-\frac{55}{4}\nu+\frac{3}{4}\nu^2+2\gamma_{\rm AB}-5\gamma_{\rm AB}\nu\Big)\tilde{p}_e^4
		\nonumber\\
		&+\Big(4+\nu-\frac{17}{2}\nu^2+4\gamma_{\rm AB}-7\gamma_{\rm AB}\nu\Big)\tilde{p}_e^2\tilde{p}_{r, e}^2\Bigg]
		\nonumber\\
		&+\frac{G_{\rm AB}^2M^2}{r_e^2}\Bigg[\Big(5+17\nu-\frac{3}{4}\nu^2+\gamma_{\rm AB}[3+7\nu]-\frac{7}{4}\gamma_{\rm AB}^2\nonumber\\
		&+2\langle\beta\rangle-\beta_+\Big)\tilde{p}_e^2+\Big(7+\frac{29}{2}\nu+\frac{9}{4}\nu^2+\gamma_{\rm AB}[10+7\nu]+3\gamma_{\rm AB}^2\nonumber\\
		&+2\langle\beta\rangle[\nu-2]+2\beta_+\nu+\langle\delta\rangle\Big)\tilde{p}_{r, e}^2\Bigg]\Bigg\}+\mathcal{O}\left(\frac{1}{c^5}\right),\nonumber\\
		\frac{1}{r_h} &= \frac{1}{r_e}+ \frac{1}{c^2}\Bigg[-\frac{\nu}{2r_e}\tilde{p}_e^2+\frac{G_{\rm AB}M}{r_e^2}\left(1+\gamma_{\rm AB}+\frac{\nu}{2}\right)\nonumber \\
		&\quad-\frac{\nu}{ r_e} \tilde{p}_{r, e}^2\Bigg]+\frac{1}{c^4}\Bigg[(\nu+3\nu^2)\frac{\tilde{p}_{e}^4}{8r_e}+\nu\frac{\tilde{p}_{e}^2\tilde{p}_{r, e}^2}{2r_e}+3\nu^2\frac{\tilde{p}_{r, e}^4}{2r_e}\nonumber\\
		&-\nu\frac{G_{\rm AB}M}{8r_e^2}\Big([-39+5\nu-32\gamma_{\rm AB}]\tilde{p}_{r, e}^2+[7+7\nu+4\gamma_{\rm AB}]\tilde{p}_{e}^2\Big)\nonumber\\
		&+\frac{G_{\rm AB}^2M^2}{r_e^3}\Big(1-\frac{15}{4}\nu+\frac{\nu^2}{2}+\frac{[5-4\nu]}{2}\gamma_{\rm AB}+\frac{11}{8}\gamma_{\rm AB}^2\nonumber\\
		&+\langle\beta\rangle+\beta_+\nu-\frac{\langle\delta\rangle}{2}\Big)\Bigg]+\mathcal{O}\left(\frac{1}{c^5}\right).
	\end{align}
	We use the above derivations of the coordinate transformation from harmonic to EOB coordinate system to express the energy and angular momentum flux in EOB coordinates \cite{Bernard:2022noq,Lange:2017wki,Lange:2018pyp, Jain:2024lie}. The energy flux in EOB coordinates is expressed as
	\begin{align}
		\phi_E^e = G_{\rm AB} c^3 \phi_E~,
	\end{align}
where $\phi_E^e$ is the energy flux in EOB coordinate system and $\phi_E$ is the physical energy flux expressed in harmonic coordinates.  Similarly, the angular momentum flux in EOB coordinates can be expressed as
	\begin{align}
		\phi_J^e =  c^3 \frac{\phi_J}{M}~,
	\end{align}
where $\phi_J^e$ is the angular momentum flux in EOB coordinate system and $\phi_J$ is the physical angular momentum flux expressed in harmonic coordinates. The explicit expression of EOB coordinate energy and angular momentum flux up to  2PN is given in the supplementary file attached.

		\section{Radiation-Reaction}
		\label{sec:Five}

			In this section, we use the EOB coordinate energy and angular momentum flux to derive the generic orbit radiation reaction force within EOB formalism for non-spinning systems.
			Since from now on we will only work in EOB coordinates, we will drop the $e$ subscript and tilde to define the dimensionless variables and denote EOB phase space coordinates with $(r,\varphi,p_r,p_\varphi)$.

			Now, let us remember that there are two contributions to the fluxes in scalar-tensor theory, i.e. a scalar contribution in addition to the usual tensorial one. In this work, we will compute the radiation reaction force due to both the scalar and tensorial contributions up to 1.5PN order. 

Numerically, in most regions of the parameter space the scalar fluxes are small contributions to the total energy or angular momentum fluxes.


Radiation reaction in the EOB framework was first computed in Ref.~\cite{Buonanno:2000ef} for quasi-circular inspirals in GR.
The arbitrary gauge freedom discussed in  Refs.~\cite{Iyer:1993xi, Iyer:1995rn} in the radiation reaction force was utilised in~\cite{Buonanno:2000ef} to find a freedom
in relating the angular ($\mathcal{F}_{\phi}$) and radial component ($\mathcal{F}_r$) of the radiation reaction force.
However, as shown above, there is no gauge freedom at 1.5PN order in scalar-tensor theories which implies that the quasi-circular orbit relations of GR do not hold in ST theories at 1.5PN order.

GR relations were later used to compute the generic orbit radiation reaction force in Ref.~\cite{Bini:2012ji} (see also \cite{Khalil:2021txt}).
We here start with the approach of Ref.~\cite{Khalil:2021txt} in GR to derive radiation reaction (RR) force that satisfy 
\begin{align}
	\label{eq:conditionGauge}
	\mathcal{F}_{\phi}^{\rm qc} &= -\frac{\Phi_E^{\rm qc}}{\Omega}~,\nonumber\\
	\mathcal{F}_r^{\rm qc} &= \mathcal{F}_{\phi}^{\rm qc}\frac{p_r}{p_{\phi}} = -\frac{\Phi_E^{\rm qc}p_r}{\Omega p_{\phi}} ~,
\end{align}
for circular orbits, where $\Omega$ is the orbital frequency.
This formulation then allows the generic orbit force to be factorised as
\begin{align}
	\mathcal{F}_r = \mathcal{F}_r^{\rm qc}\mathcal{F}_r^{\rm ecc}~,\nonumber \\
	\mathcal{F}_{\phi} = \mathcal{F}_{\phi}^{\rm qc}\mathcal{F}_{\phi}^{\rm ecc}~,
\end{align} 
where $\mathcal{F}_X^{\rm ecc}$ now denote the eccentric fluxes.

To derive the radiation reaction force in the EOB coordinates using the above balance equations, we use the energy flux and angular momentum flux in the EOB coordinates and write a generic ansatz with the unknown coefficients for the Schott terms. Then, we calculate the unknown coefficients in the Schott terms by requiring that the force be reduced to the radiation reaction force in the limit of quasi-circular orbits,
\begin{align}
	\label{eq:condrr}
	\mathcal{F}_{\phi}&=-\Phi_J+\mathcal{O}\left(p_r^2\right)+\mathcal{O}\left(\dot{p}_r\right),\nonumber\\
	\frac{\mathcal{F}_{r}p_{\phi}}{\mathcal{F}_{\phi}p_{r}}&=1+\mathcal{O}\left(p_r^2\right),
\end{align}
where the radial momenta $p_r$ and its first derivative $\dot{p}_r$ are zero for circular orbits. 
As discussed previously, we will show that Eqs.~\eqref{eq:condrr}, which are valid in General Relativity (GR), cannot be upheld in scalar-tensor (ST) theories and therefore must be relaxed.

\subsection{Ansatz for Schott terms}

The general ansatz for the Schott terms is dictated by the fact that these terms enter due to the interaction with the radiation field, i.e., these are radiation effects. Therefore, we follow the postulates of constructing the radiation reaction acceleration given in Eq.~\eqref{eqn:rr-general} and additionally requiring that the angular momentum remains a pseudo-vector to construct the general ansatz. 
This implies that the Schott energy and angular momentum must be proportional to the odd powers in $p_r$, $m/r$ for the non-linear effects and $\nu^2$ for vanishing radiation effects in the test mass limit. 

However, before expressing the	general form of the ansatz for Schott terms, we recall that due to the presence of the scalar field (and hence, the dipolar radiation due to the scalar field) in scalar-tensor theories, the contribution to flux starts one PN order before their GR equivalent, i.e., at -1PN order. 
In order to account of this information for the construction of general ansatz, we first explicitly compute both the energy and the angular momentum Schott terms at -1PN order. 
To do this, we consider an analogy with the dipolar radiation due to an electric charge in electrodynamics.
The radiation-reaction force (Abraham-Lorentz Force) for a slowly moving point particle with charge $e$ is \cite{landau1975classical},
\begin{equation}
	F_{\rm rad}^{\rm ED} =  \frac{2 e^2}{3 c^3} \dot{a}~,
\end{equation}
where $a$ is the acceleration and the superscript `ED' is to specify electrodynamics. In terms of the multipole moments the above equation can then be expressed as
\begin{equation}
	F_{i }^{\rm rad, ED} =  \frac{2e}{3 c^3} I^{(3)}_i~,
\end{equation}
where $I^{(3)}_i$ is the electric dipole moment. This implies that the radiation reaction force for a point particle moving in a scalar field must also be proportional to the third-time derivative of the scalar dipole moment,
\begin{equation}
	\label{eq:FradScalar}
	F_{i }^{\rm rad} = C_s I^{s^{(3)}}_i~,
\end{equation}
where $I^s_i  =\frac{ 4 \alpha_{\rm AB} (\alpha_B^0-\alpha_A^0) \zeta M \nu}{(1-\zeta)} r_i = S_{\rm AB} r_i$ is the scalar dipole moment and $C_s$ is some constant of proportionality. Therefore, the energy and angular momentum radiated from the system due to the scalar dipolar radiation are,
\begin{align}
	\dot{E}_{\rm system} &= F_i^{\rm rad} v^{i}~,\nonumber\\
	\dot{J}_{\rm system} &= \epsilon^{ijk} r_{i} F_{j}^{\rm rad}~, 
\end{align}
where $r_{i}$ and $v_{i}$ are the position and the velocity of the particle, respectively. Using Eq.~\eqref{eq:FradScalar} in the above equations, we obtain
\begin{align}
	\label{eq:EnradScalar}
	\dot{E}_{\rm system} &= C_s I^{s^{(3)}}_i v^{i} \nonumber\\
	& = \frac{C_{s}}{S_{\rm AB}} I^{s^{(3)}}_i \dot{I}_s^i~,
\end{align}
and 
\begin{align}
	\label{eq:JradScalar}
	\dot{J}_{\rm system} &= \epsilon^{ijk}C_s I^{s^{(3)}}_j r_{i} \nonumber\\
	& = \frac{C_{s}}{S_{\rm AB}} \epsilon^{ijk} I^{s}_i I^{s^{(3)}}_j ~.
\end{align}
The radiated energy given in Eq.~\eqref{eq:EnradScalar} can be expressed in terms of the total time derivative term as
\begin{align}
	\dot{E}_{\rm system}  & = \frac{C_{s}}{S_{\rm AB}} \left[\frac{d}{dt}\left(\ddot{I}^s_i \dot{I}^s_i\right)-\ddot{I}^s_i \ddot{I}^s_i\right]~.
\end{align}
Now, using the above equation and the balance law equation given in Eq.~\eqref{eq:EandJbalancelaw}, we find that the leading order energy Schott term, $E_{(\rm schott)}$,  due to the scalar dipolar radiation is
\begin{align}
	\label{eq:ESchottLO}
	E_{(\rm schott)} =\frac{C_{s}}{S_{\rm AB}} \ddot{I}^s_i \dot{I}^s_i~.
\end{align}

Similarly, for the radiated angular momentum given in Eq.~\eqref{eq:JradScalar} can be expressed as,
\begin{align}
	\dot{J}_{\rm system}  & = \frac{C_{s}}{S_{\rm AB}}\epsilon^{ijk}\left[\frac{d}{dt}\left( I^{s}_i \ddot{I}^{s}_j \right)-\dot{I}^s_i \ddot{I}^s_j\right]~,
\end{align}
and the leading order angular momentum Schott term, $J_{(\rm schott)}$,  due to the scalar dipolar radiation is,
\begin{align}
	\label{eq:JSchottLO}
	J_{(\rm schott)} &= \frac{C_{s}}{S_{\rm AB}}\epsilon^{ijk} I^{s}_i \ddot{I}^{s}_j~.
\end{align}
Using the definition of the scalar dipole moment and the Newtonian order equations of motion in Eqs.~ \eqref{eq:ESchottLO} and \eqref{eq:JSchottLO}, we obtain that for the dipolar radiation there is no angular momentum contribution to the Schott term, i.e. $J_{(\rm schott)} =0$, whereas the energy Schott term is non-zero.

Furthermore, we note that the energy and the angular momentum flux at 0.5PN order (the radiation-reaction square effects) is a total derivative term, implying that these can be used to define a new conserved quantity. Therefore, these terms must not contribute in the radiation-reaction force. 

After considering the above two observations, the ansatz for the Schott terms we use is,
\begin{align}
	\text{J}_{\rm Schott}&=\frac{\nu ^2 p_{\phi} p_{r}}{r^2} \Bigg[\frac{\xi _1}{c^2}+ \frac{1}{c^3}\frac{\xi_2}{p_r r}\Bigg]\,,\\
	E_{\rm Schott}&=\frac{\nu ^2 p_{r}}{r^2} \Bigg[\psi_0 + \frac{1}{c^2}\left(\psi_1 p_{r}^2+\psi _2 p^2+\frac{\psi _3}{r}\right)\nonumber\\
	&+\frac{1}{c^3}\frac{\psi_4}{p_r r^2}\Bigg]\,.
\end{align}

\subsection{Generic orbit RR force}

Using the fluxes and Schott terms, the radiation reaction force is calculated using the balance equation and the radiation reaction condition in the circular-orbit limit.

\subsubsection{-1PN order}

As mentioned previously, scalar-tensor contributions start one PN order before their GR equivalents.
We also noted how there is no gauge freedom in the radiation reaction term at 1.5PN order between Harmonic coordinate system and ADM coordinate system. 

At this order, the RR fluxes read
\begin{align}
	\mathcal{F}_\varphi &= -\frac{4}{3} \nu^2 \zeta^{\prime} \mathcal{S}_{-}^2\, p_{\varphi}\, u^3 + O\left(\frac{1}{c^2}\right)\,, \nonumber \\
	\mathcal{F}_r &= \frac{4}{3} \nu^2 u^3 \Bigg[
	\frac{3}{2} \psi_0\, p_{r} + \left(\zeta^{\prime} \mathcal{S}_{-}^2 - \frac{3}{4}\psi_0\right) \frac{p_{\varphi}^2 u^2}{p_{r}} \nonumber \\
	&\quad - \left(\zeta^{\prime} \mathcal{S}_{-}^2 - \frac{3}{4}\psi_0\right) \frac{u}{p_{r}} \Bigg] + O\left(\frac{1}{c^2}\right)\,,
\end{align}
where we defined $u \equiv 1/r$ and $p_{\varphi}$ is such that $p^2 = p_{r}^2 + p_{\varphi}^2 u^2$.

As mentioned previously,  the unknown parameters entering the ansatz of the Schott terms are determined by satisfying the circular orbit limit, which requires that the radial flux in the above equation must not contain negative powers of the radial momentum $p_{r}$. This leads to the solution,
\begin{equation}
	\psi_0 = \frac{4}{3} \zeta^{\prime} \mathcal{S}_{-}^2\,,
\end{equation}
so that
\begin{equation}
	\mathcal{F}_r = \frac{8}{3} \nu^2 \zeta^{\prime} \mathcal{S}_{-}^2 p_{r} u^3  + O\left(\frac{1}{c^2}\right)\,.
\end{equation}

At this level, it is easy to see we cannot impose $(\mathcal{F}_{r}p_{\phi})/(\mathcal{F}_{\phi}p_{r})=1$. Instead we get
\begin{equation}
	\frac{\mathcal{F}_{r}p_{\phi}}{\mathcal{F}_{\phi}p_{r}}=-2\,.
\end{equation}

From here on, we will use the conditions 
\begin{align}
	\label{eq:condSTrr}
	\mathcal{F}_{\phi}&=-\Phi_J+\mathcal{O}\left(p_r^2\right)+\mathcal{O}\left(\dot{p}_r\right),\nonumber\\
	\frac{\mathcal{F}_{r}p_{\phi}}{\mathcal{F}_{\phi}p_{r}}&=1+\mathcal{O}\left(p_r^2\right),
\end{align}
in the limit of quasi-circular orbits for scalar-tensor theories.
 
\subsubsection{Newtonian Order}
At Newtonian order, we follow the same steps as above, i.e. impose that the radial momentum contains only positive powers of $p_{r}$, in addition to Eq.~\eqref{eq:condSTrr} to determine all four free parameters, as
\begin{align}
	\xi_1 &= \frac{8}{15}\left(2+\gamma_{\rm AB}\right) +\frac{4}{15} \zeta^\prime \Bigg[ 14\mathcal{S}_+^2 -  \mathcal{S}_+(3 \mathcal{S}_-  X_{\rm AB}  + 80 \theta_+)
	\nonumber\\
	& + 10 \mathcal{S}_-( \theta_- + X_{\rm AB} \theta_+)-\mathcal{S}_-^2 (6-5\langle\bar{\beta}\rangle + 5 \gamma_{\rm AB} + \nu)
	\nonumber\\
	&+120 \theta_+^2 \Bigg]\,,\nonumber\\
	\psi_1 &= -\frac{4}{3}\left(2+\gamma_{\rm AB}\right) -\frac{2}{3} \zeta^\prime \Bigg[6\mathcal{S}_{+}^2 -32 \theta_+ \mathcal{S}_+ + 8 \theta_- \mathcal{S}_-
	\nonumber\\
	&-\mathcal{S}_-^2(3-2\langle\bar{\beta}\rangle+2\bar{\gamma}_{\rm AB})+ 40 \theta_+^2 \Bigg]\,,\nonumber\\
	\psi_2 &= \frac{32}{15}\left(2+\gamma_{\rm AB}\right) + \frac{2}{15} \Bigg[32 \mathcal{S}_+^2 - 160  \theta_+ \mathcal{S}_+ + 40  \theta_- \mathcal{S}_-
	\nonumber\\
	&-\mathcal{S}_-^2 (2-10\langle\bar{\beta}\rangle+15\nu) + 240 \theta_+^2\Bigg]\,,\nonumber\\
	\psi_3 &= -\frac{4}{15}\mathcal{S}_- \zeta'\Bigg[4 \mathcal{S}_+ X_{\rm AB} + \mathcal{S}_-\Bigg(31 + 10\langle\bar{\beta}\rangle 
	\nonumber\\
	&+ 10 \bar{\gamma}_{\rm AB}-7\nu \Bigg)\Bigg] \,.
\end{align}

At this order, the final forms for the radiation reaction forces are 
\begin{align}
	\mathcal{F}_\varphi &= -\frac{4}{3} \nu^2 \zeta^{\prime} \mathcal{S}_{-}^2\, p_{\varphi}\, u^3 + \frac{1}{c^2} \mathcal{F}_\varphi^{\rm N} + O\left(\frac{1}{c^3}\right) \,, \nonumber \\
	\mathcal{F}_r &= \frac{8}{3} \nu^2 \zeta^{\prime} \mathcal{S}_{-}^2\, p_{r}\, u^3 + \frac{1}{c^2} \mathcal{F}_r^{\rm N} + O\left(\frac{1}{c^3}\right) \,,
\end{align}
with
\begin{align}
	\mathcal{F}_\varphi^{\rm N} &= \frac{2}{15}\nu^2 p_{\varphi} u^3 \Bigg\{\Bigg[-16(2+\bar{\gamma}_{\rm AB})- \zeta^\prime \big(32 \mathcal{S}_+^2 - 160 \theta_+ \mathcal{S}_+ 
	\nonumber\\
	&+ 40 \theta_- \mathcal{S}_- - \mathcal{S}_-^2(2-10\langle\bar{\beta}\rangle+15\nu) + 240 \theta_+^2\big) \Bigg] p^2
		\nonumber\\
	&+15\Bigg[2(2+\bar{\gamma}_{\rm AB}) + \zeta^\prime\big(6 \mathcal{S}_+^2 -32 \theta_+ \mathcal{S}_+ +8 \theta_- \mathcal{S}_- 
				\nonumber\\
	&+ \mathcal{S}_-^2(-3 + 2\langle\bar{\beta}\rangle -2 \bar{\gamma}_{\rm AB})+48 \theta_+^2\big)\Bigg]p_r^2-\Bigg[8(2+\bar{\gamma}_{\rm AB})
		\nonumber\\
	&-\zeta^\prime\big(24 \mathcal{S}_+^2 - 160 \theta_+\mathcal{S}_+ + 40 X_{\rm AB} \theta_+ \mathcal{S}_- + \mathcal{S}_-^2(41+30 \langle\bar{\beta}\rangle-10\nu) 	
	\nonumber\\
	&+ 240 \theta_+^2\big)\Bigg]u \Bigg\}  \,, \nonumber \\
	\mathcal{F}_r^{\rm N} &= \frac{4}{15}\nu^2 p_r u^3 \Bigg\{\Bigg[16(2+\bar{\gamma}_{\rm AB})+ \zeta^\prime \big(32 \mathcal{S}_+^2 - 160 \theta_+ \mathcal{S}_+ + 40 \theta_- \mathcal{S}_- 
	\nonumber\\
	&- \mathcal{S}_-^2(2-10\langle\bar{\beta}\rangle+15\nu) + 240 \theta_+^2\big) \Bigg] p^2 - 5\Bigg[2(2+\bar{\gamma}_{\rm AB}) + \zeta^\prime\big(6 \mathcal{S}_+^2 
			\nonumber\\
	&-32 \theta_+ \mathcal{S}_+ +8 \theta_- \mathcal{S}_- + \mathcal{S}_-^2(-3 + 2\langle\bar{\beta}\rangle -2 \bar{\gamma}_{\rm AB})+48 \theta_+^2\big)\Bigg]p_r^2 
	\nonumber\\
	&+\Bigg[8(2+\bar{\gamma}_{\rm AB})-\zeta^\prime\big(24 \mathcal{S}_+^2 - 160 \theta_+\mathcal{S}_+ + 40 X_{\rm AB} \theta_+ \mathcal{S}_- 
	\nonumber\\
	&+ \mathcal{S}_-^2(41+30 \langle\bar{\beta}\rangle-10\nu) + 240 \theta_+^2\big)\Bigg]u\Bigg\} \,.
	\label{eq:NFrandp}
\end{align}

\subsubsection{0.5PN}
Again following the same steps as previously to determine the unknowns entering at 0.5PN order. However, instead of requiring that $1/p_r$ term in $\mathcal{F}_r$ vanishes , we now impose that the $\mathcal{O}(p_r^0)$ vanishes (as $\mathcal{F}_r$ vanishes in the quasi circular orbit limit).  This is because, at the 0.5PN order, the leading contribution is of order $\mathcal{O}(p_r^0)$ unlike -1PN and Newtonian order where such terms disappear and leading contribution is $\mathcal{O}(1/p_r)$.  This along with the condition $(\mathcal{F}_{r}p_{\phi})/(\mathcal{F}_{\phi}p_{r})=-2$, leads to the solution,
\begin{align}
\psi_4 & =  -\frac{2}{9} \mathcal{S}_-^2 \zeta^\prime\Big[4\mathcal{S}_- (\mathcal{S}_- + X_{\rm AB}\mathcal{S}_+)\zeta^\prime -\bar{\gamma}_{\rm AB}\Big] \,,
\nonumber\\
\xi_2 & = -\frac{2}{9} \mathcal{S}_-^2 \zeta^\prime\Big[4\mathcal{S}_- (\mathcal{S}_- + X_{\rm AB}\mathcal{S}_+)\zeta^\prime -\bar{\gamma}_{\rm AB}\Big]\,.
\end{align} 

At this order, the radiation reaction force vanishes, consistent with the fact that at 0.5PN order the energy and the angular momentum flux are total derivatives, implying that these can be used to define a new conserved quantity.  Hence, the radiation-reaction force up to 0.5PN reads,
\begin{align}
	\mathcal{F}_\varphi &= -\frac{4}{3} \nu^2 \zeta^{\prime} \mathcal{S}_{-}^2\, p_{\varphi}\, u^3 + \frac{1}{c^2} \mathcal{F}_\varphi^{\rm N} + O\left(\frac{1}{c^4}\right) \,, \nonumber \\
	\mathcal{F}_r &= \frac{8}{3} \nu^2 \zeta^{\prime} \mathcal{S}_{-}^2\, p_{r}\, u^3 + \frac{1}{c^2} \mathcal{F}_r^{\rm N} + O\left(\frac{1}{c^4}\right) \,,
\end{align}
with $\mathcal{F}_\varphi^{\rm N}$ and $\mathcal{F}_r^{\rm N}$ are given in Eq.~\eqref{eq:NFrandp}.

\section{Discussions}
\label{sec:Six}

Recently there has been a significant effort to extend EOB waveform models to incorporate eccentricity effects in GR~\cite{Bini:2012ji,Hinderer:2017jcs,Gamboa:2024hli,Albanesi:2025txj} 
In this paper, we extend the quasi-circular orbit approximation results of the radiation reaction force in massless scalar-tensor theories to eccentric orbits for nonspinning black hole binaries within the EOB framework. 

In particular, we first derive the coordinate transformations required to map the energy and angular momentum fluxes from harmonic coordinates to EOB coordinates up to 2PN order by deriving the transformation from harmonic to ADM coordinate system and from ADM to EOB coordinate system. As in scalar-tensor theories, the radiation reaction effects start 1PN order before (1.5PN order) in contrast to GR, we also discuss the coordinate transformation from harmonic coordinates to ADM-like coordinates by investigating the gauge freedom in the radiation reaction term following Ref.~\cite{Iyer:1993xi, Iyer:1995rn}. Unlike in GR \cite{Iyer:1993xi, Iyer:1995rn}, we find that there is no gauge freedom in the radiation reaction effects at 1.5PN order for scalar-tensor theories, implying that there is no coordinate transformation between the harmonic and ADM-like coordinate systems at 1.5PN order.  We also verified that the coordinate transformation between harmonic, ADM and EOB coordinates systems are in agreement with existing PN-based results in the GR-limit.

We hence made use of 2PN-order coordinate transformations between harmonic, ADM and EOB frames to translate the harmonic coordinate energy and angular momentum fluxes derived in Refs.~\cite{Bernard:2022noq, Jain:2024lie} into the EOB coordinate system. Next, we construct the general ansatz of the Schott terms by considering an analogy of the scalar dipolar radiation with the dipolar radiation due to an electric charge in electrodynamics and found that there is no angular momentum contribution to the Schott term, i.e. $J_{\rm (schott)} = 0$ at the leading order (-1PN order with respect to GR) in scalar-tensor theories. 
Using these results, we derive the radiation reaction force for generic orbits up to relative 1.5PN order, i.e. -1PN, Newtonian and 0.5PN order, where we find that at the 0.5PN order the energy and the angular momentum flux are total derivatives, suggesting that the radiation reaction force vanishes at 0.5PN order.

The results presented in this paper will be useful when building accurate GW models that take into account multiple effects, such as eccentricity and ST corrections.
Such models could then be employed in tests of GR in order to avoid possible biases due to the omission of one such effect.

\section*{Acknowledgments}
T.~J. and P.~R. thank the hospitality and the stimulating environment of the Institut des Hautes Etudes Scientifiques.  T. J. is supported by a LabEx Junior Research Chair Fellowship. 
The present research was also partly supported by the ``\textit{2021 Balzan Prize for 
Gravitation: Physical and Astrophysical Aspects}'', awarded to Thibault Damour. 
P.~R.~acknowledges support from the Italian Ministry of University and Research (MUR) via the PRIN 2022ZHYFA2, GRavitational wavEform models for coalescing compAct binaries with eccenTricity (GREAT) and through the program
“Dipartimenti di Eccellenza 2018-2022” (Grant SUPER-C).
P.~R.~also acknowledges support from “Fondo di Ricerca d’Ateneo” of the University of Perugia. 
	
\bibliographystyle{apsrev4-1}
\bibliography{refs.bib, local.bib}

\begin{thebibliography}{56}%
\makeatletter
\providecommand \@ifxundefined [1]{%
 \@ifx{#1\undefined}
}%
\providecommand \@ifnum [1]{%
 \ifnum #1\expandafter \@firstoftwo
 \else \expandafter \@secondoftwo
 \fi
}%
\providecommand \@ifx [1]{%
 \ifx #1\expandafter \@firstoftwo
 \else \expandafter \@secondoftwo
 \fi
}%
\providecommand \natexlab [1]{#1}%
\providecommand \enquote  [1]{``#1''}%
\providecommand \bibnamefont  [1]{#1}%
\providecommand \bibfnamefont [1]{#1}%
\providecommand \citenamefont [1]{#1}%
\providecommand \href@noop [0]{\@secondoftwo}%
\providecommand \href [0]{\begingroup \@sanitize@url \@href}%
\providecommand \@href[1]{\@@startlink{#1}\@@href}%
\providecommand \@@href[1]{\endgroup#1\@@endlink}%
\providecommand \@sanitize@url [0]{\catcode `\\12\catcode `\$12\catcode
  `\&12\catcode `\#12\catcode `\^12\catcode `\_12\catcode `\%12\relax}%
\providecommand \@@startlink[1]{}%
\providecommand \@@endlink[0]{}%
\providecommand \url  [0]{\begingroup\@sanitize@url \@url }%
\providecommand \@url [1]{\endgroup\@href {#1}{\urlprefix }}%
\providecommand \urlprefix  [0]{URL }%
\providecommand \Eprint [0]{\href }%
\providecommand \doibase [0]{http://dx.doi.org/}%
\providecommand \selectlanguage [0]{\@gobble}%
\providecommand \bibinfo  [0]{\@secondoftwo}%
\providecommand \bibfield  [0]{\@secondoftwo}%
\providecommand \translation [1]{[#1]}%
\providecommand \BibitemOpen [0]{}%
\providecommand \bibitemStop [0]{}%
\providecommand \bibitemNoStop [0]{.\EOS\space}%
\providecommand \EOS [0]{\spacefactor3000\relax}%
\providecommand \BibitemShut  [1]{\csname bibitem#1\endcsname}%
\let\auto@bib@innerbib\@empty
\bibitem [{\citenamefont {Abbott}\ \emph {et~al.}(2016)\citenamefont {Abbott}
  \emph {et~al.}}]{LIGOScientific:2016aoc}%
  \BibitemOpen
  \bibfield  {author} {\bibinfo {author} {\bibfnamefont {B.~P.}\ \bibnamefont
  {Abbott}} \emph {et~al.} (\bibinfo {collaboration} {LIGO Scientific,
  Virgo}),\ }\href {\doibase 10.1103/PhysRevLett.116.061102} {\bibfield
  {journal} {\bibinfo  {journal} {Phys. Rev. Lett.}\ }\textbf {\bibinfo
  {volume} {116}},\ \bibinfo {pages} {061102} (\bibinfo {year} {2016})},\
  \Eprint {http://arxiv.org/abs/1602.03837} {arXiv:1602.03837 [gr-qc]}
  \BibitemShut {NoStop}%
\bibitem [{\citenamefont {Reitze}\ \emph {et~al.}(2019)\citenamefont {Reitze}
  \emph {et~al.}}]{Reitze:2019iox}%
  \BibitemOpen
  \bibfield  {author} {\bibinfo {author} {\bibfnamefont {D.}~\bibnamefont
  {Reitze}} \emph {et~al.},\ }\href@noop {} {\bibfield  {journal} {\bibinfo
  {journal} {Bull. Am. Astron. Soc.}\ }\textbf {\bibinfo {volume} {51}},\
  \bibinfo {pages} {035} (\bibinfo {year} {2019})},\ \Eprint
  {http://arxiv.org/abs/1907.04833} {arXiv:1907.04833 [astro-ph.IM]}
  \BibitemShut {NoStop}%
\bibitem [{\citenamefont {Punturo}\ \emph {et~al.}(2010)\citenamefont
  {Punturo}, \citenamefont {Abernathy}, \citenamefont {Acernese}, \citenamefont
  {Allen}, \citenamefont {Andersson} \emph {et~al.}}]{Punturo:2010zz}%
  \BibitemOpen
  \bibfield  {author} {\bibinfo {author} {\bibfnamefont {M.}~\bibnamefont
  {Punturo}}, \bibinfo {author} {\bibfnamefont {M.}~\bibnamefont {Abernathy}},
  \bibinfo {author} {\bibfnamefont {F.}~\bibnamefont {Acernese}}, \bibinfo
  {author} {\bibfnamefont {B.}~\bibnamefont {Allen}}, \bibinfo {author}
  {\bibfnamefont {N.}~\bibnamefont {Andersson}},  \emph {et~al.},\ }\href
  {\doibase 10.1088/0264-9381/27/19/194002} {\bibfield  {journal} {\bibinfo
  {journal} {Class.Quant.Grav.}\ }\textbf {\bibinfo {volume} {27}},\ \bibinfo
  {pages} {194002} (\bibinfo {year} {2010})}\BibitemShut {NoStop}%
\bibitem [{\citenamefont {Amaro-Seoane}(2018)}]{Amaro-Seoane:2018gbb}%
  \BibitemOpen
  \bibfield  {author} {\bibinfo {author} {\bibfnamefont {P.}~\bibnamefont
  {Amaro-Seoane}},\ }\href {\doibase 10.1103/PhysRevD.98.063018} {\bibfield
  {journal} {\bibinfo  {journal} {Phys. Rev. D}\ }\textbf {\bibinfo {volume}
  {98}},\ \bibinfo {pages} {063018} (\bibinfo {year} {2018})},\ \Eprint
  {http://arxiv.org/abs/1807.03824} {arXiv:1807.03824 [astro-ph.HE]}
  \BibitemShut {NoStop}%
\bibitem [{\citenamefont {Maggiore}\ \emph {et~al.}(2020)\citenamefont
  {Maggiore} \emph {et~al.}}]{Maggiore:2019uih}%
  \BibitemOpen
  \bibfield  {author} {\bibinfo {author} {\bibfnamefont {M.}~\bibnamefont
  {Maggiore}} \emph {et~al.},\ }\href {\doibase 10.1088/1475-7516/2020/03/050}
  {\bibfield  {journal} {\bibinfo  {journal} {JCAP}\ }\textbf {\bibinfo
  {volume} {03}},\ \bibinfo {pages} {050} (\bibinfo {year} {2020})},\ \Eprint
  {http://arxiv.org/abs/1912.02622} {arXiv:1912.02622 [astro-ph.CO]}
  \BibitemShut {NoStop}%
\bibitem [{\citenamefont {Evans}\ \emph {et~al.}(2021)\citenamefont {Evans}
  \emph {et~al.}}]{Evans:2021gyd}%
  \BibitemOpen
  \bibfield  {author} {\bibinfo {author} {\bibfnamefont {M.}~\bibnamefont
  {Evans}} \emph {et~al.},\ }\href@noop {} {\  (\bibinfo {year} {2021})},\
  \Eprint {http://arxiv.org/abs/2109.09882} {arXiv:2109.09882 [astro-ph.IM]}
  \BibitemShut {NoStop}%
\bibitem [{\citenamefont {Damour}\ and\ \citenamefont
  {Esposito-Farese}(1996{\natexlab{a}})}]{Damour:1996ke}%
  \BibitemOpen
  \bibfield  {author} {\bibinfo {author} {\bibfnamefont {T.}~\bibnamefont
  {Damour}}\ and\ \bibinfo {author} {\bibfnamefont {G.}~\bibnamefont
  {Esposito-Farese}},\ }\href {\doibase 10.1103/PhysRevD.54.1474} {\bibfield
  {journal} {\bibinfo  {journal} {Phys. Rev. D}\ }\textbf {\bibinfo {volume}
  {54}},\ \bibinfo {pages} {1474} (\bibinfo {year} {1996}{\natexlab{a}})},\
  \Eprint {http://arxiv.org/abs/gr-qc/9602056} {arXiv:gr-qc/9602056}
  \BibitemShut {NoStop}%
\bibitem [{\citenamefont {Freire}\ \emph {et~al.}(2012)\citenamefont {Freire},
  \citenamefont {Wex}, \citenamefont {Esposito-Farese}, \citenamefont
  {Verbiest}, \citenamefont {Bailes}, \citenamefont {Jacoby}, \citenamefont
  {Kramer}, \citenamefont {Stairs}, \citenamefont {Antoniadis},\ and\
  \citenamefont {Janssen}}]{Freire:2012mg}%
  \BibitemOpen
  \bibfield  {author} {\bibinfo {author} {\bibfnamefont {P.~C.~C.}\
  \bibnamefont {Freire}}, \bibinfo {author} {\bibfnamefont {N.}~\bibnamefont
  {Wex}}, \bibinfo {author} {\bibfnamefont {G.}~\bibnamefont
  {Esposito-Farese}}, \bibinfo {author} {\bibfnamefont {J.~P.~W.}\ \bibnamefont
  {Verbiest}}, \bibinfo {author} {\bibfnamefont {M.}~\bibnamefont {Bailes}},
  \bibinfo {author} {\bibfnamefont {B.~A.}\ \bibnamefont {Jacoby}}, \bibinfo
  {author} {\bibfnamefont {M.}~\bibnamefont {Kramer}}, \bibinfo {author}
  {\bibfnamefont {I.~H.}\ \bibnamefont {Stairs}}, \bibinfo {author}
  {\bibfnamefont {J.}~\bibnamefont {Antoniadis}}, \ and\ \bibinfo {author}
  {\bibfnamefont {G.~H.}\ \bibnamefont {Janssen}},\ }\href {\doibase
  10.1111/j.1365-2966.2012.21253.x} {\bibfield  {journal} {\bibinfo  {journal}
  {Mon. Not. Roy. Astron. Soc.}\ }\textbf {\bibinfo {volume} {423}},\ \bibinfo
  {pages} {3328} (\bibinfo {year} {2012})},\ \Eprint
  {http://arxiv.org/abs/1205.1450} {arXiv:1205.1450 [astro-ph.GA]} \BibitemShut
  {NoStop}%
\bibitem [{\citenamefont {Shao}\ \emph {et~al.}(2017)\citenamefont {Shao},
  \citenamefont {Sennett}, \citenamefont {Buonanno}, \citenamefont {Kramer},\
  and\ \citenamefont {Wex}}]{Shao:2017gwu}%
  \BibitemOpen
  \bibfield  {author} {\bibinfo {author} {\bibfnamefont {L.}~\bibnamefont
  {Shao}}, \bibinfo {author} {\bibfnamefont {N.}~\bibnamefont {Sennett}},
  \bibinfo {author} {\bibfnamefont {A.}~\bibnamefont {Buonanno}}, \bibinfo
  {author} {\bibfnamefont {M.}~\bibnamefont {Kramer}}, \ and\ \bibinfo {author}
  {\bibfnamefont {N.}~\bibnamefont {Wex}},\ }\href {\doibase
  10.1103/PhysRevX.7.041025} {\bibfield  {journal} {\bibinfo  {journal} {Phys.
  Rev. X}\ }\textbf {\bibinfo {volume} {7}},\ \bibinfo {pages} {041025}
  (\bibinfo {year} {2017})},\ \Eprint {http://arxiv.org/abs/1704.07561}
  {arXiv:1704.07561 [gr-qc]} \BibitemShut {NoStop}%
\bibitem [{\citenamefont {Kramer}\ \emph {et~al.}(2021)\citenamefont {Kramer}
  \emph {et~al.}}]{Kramer:2021jcw}%
  \BibitemOpen
  \bibfield  {author} {\bibinfo {author} {\bibfnamefont {M.}~\bibnamefont
  {Kramer}} \emph {et~al.},\ }\href {\doibase 10.1103/PhysRevX.11.041050}
  {\bibfield  {journal} {\bibinfo  {journal} {Phys. Rev. X}\ }\textbf {\bibinfo
  {volume} {11}},\ \bibinfo {pages} {041050} (\bibinfo {year} {2021})},\
  \Eprint {http://arxiv.org/abs/2112.06795} {arXiv:2112.06795 [astro-ph.HE]}
  \BibitemShut {NoStop}%
\bibitem [{\citenamefont {Zhao}\ \emph {et~al.}(2022)\citenamefont {Zhao},
  \citenamefont {Freire}, \citenamefont {Kramer}, \citenamefont {Shao},\ and\
  \citenamefont {Wex}}]{Zhao:2022vig}%
  \BibitemOpen
  \bibfield  {author} {\bibinfo {author} {\bibfnamefont {J.}~\bibnamefont
  {Zhao}}, \bibinfo {author} {\bibfnamefont {P.~C.~C.}\ \bibnamefont {Freire}},
  \bibinfo {author} {\bibfnamefont {M.}~\bibnamefont {Kramer}}, \bibinfo
  {author} {\bibfnamefont {L.}~\bibnamefont {Shao}}, \ and\ \bibinfo {author}
  {\bibfnamefont {N.}~\bibnamefont {Wex}},\ }\href {\doibase
  10.1088/1361-6382/ac69a3} {\bibfield  {journal} {\bibinfo  {journal} {Class.
  Quant. Grav.}\ }\textbf {\bibinfo {volume} {39}},\ \bibinfo {pages} {11LT01}
  (\bibinfo {year} {2022})},\ \Eprint {http://arxiv.org/abs/2201.03771}
  {arXiv:2201.03771 [astro-ph.HE]} \BibitemShut {NoStop}%
\bibitem [{\citenamefont {Gautam}\ \emph {et~al.}(2022)\citenamefont {Gautam},
  \citenamefont {Freire}, \citenamefont {Batrakov}, \citenamefont {Kramer},
  \citenamefont {Miao}, \citenamefont {Parent},\ and\ \citenamefont
  {Zhu}}]{Gautam:2022cpb}%
  \BibitemOpen
  \bibfield  {author} {\bibinfo {author} {\bibfnamefont {T.}~\bibnamefont
  {Gautam}}, \bibinfo {author} {\bibfnamefont {P.~C.~C.}\ \bibnamefont
  {Freire}}, \bibinfo {author} {\bibfnamefont {A.}~\bibnamefont {Batrakov}},
  \bibinfo {author} {\bibfnamefont {M.}~\bibnamefont {Kramer}}, \bibinfo
  {author} {\bibfnamefont {C.~C.}\ \bibnamefont {Miao}}, \bibinfo {author}
  {\bibfnamefont {E.}~\bibnamefont {Parent}}, \ and\ \bibinfo {author}
  {\bibfnamefont {W.~W.}\ \bibnamefont {Zhu}},\ }\href {\doibase
  10.1051/0004-6361/202244699} {\bibfield  {journal} {\bibinfo  {journal}
  {Astron. Astrophys.}\ }\textbf {\bibinfo {volume} {668}},\ \bibinfo {pages}
  {A187} (\bibinfo {year} {2022})},\ \Eprint {http://arxiv.org/abs/2210.03464}
  {arXiv:2210.03464 [astro-ph.HE]} \BibitemShut {NoStop}%
\bibitem [{\citenamefont {Healy}\ \emph {et~al.}(2012)\citenamefont {Healy},
  \citenamefont {Bode}, \citenamefont {Haas}, \citenamefont {Pazos},
  \citenamefont {Laguna}, \citenamefont {Shoemaker},\ and\ \citenamefont
  {Yunes}}]{Healy:2011ef}%
  \BibitemOpen
  \bibfield  {author} {\bibinfo {author} {\bibfnamefont {J.}~\bibnamefont
  {Healy}}, \bibinfo {author} {\bibfnamefont {T.}~\bibnamefont {Bode}},
  \bibinfo {author} {\bibfnamefont {R.}~\bibnamefont {Haas}}, \bibinfo {author}
  {\bibfnamefont {E.}~\bibnamefont {Pazos}}, \bibinfo {author} {\bibfnamefont
  {P.}~\bibnamefont {Laguna}}, \bibinfo {author} {\bibfnamefont
  {D.}~\bibnamefont {Shoemaker}}, \ and\ \bibinfo {author} {\bibfnamefont
  {N.}~\bibnamefont {Yunes}},\ }\href {\doibase 10.1088/0264-9381/29/23/232002}
  {\bibfield  {journal} {\bibinfo  {journal} {Class. Quant. Grav.}\ }\textbf
  {\bibinfo {volume} {29}},\ \bibinfo {pages} {232002} (\bibinfo {year}
  {2012})},\ \Eprint {http://arxiv.org/abs/1112.3928} {arXiv:1112.3928 [gr-qc]}
  \BibitemShut {NoStop}%
\bibitem [{\citenamefont {Barausse}\ \emph {et~al.}(2013)\citenamefont
  {Barausse}, \citenamefont {Palenzuela}, \citenamefont {Ponce},\ and\
  \citenamefont {Lehner}}]{Barausse:2012da}%
  \BibitemOpen
  \bibfield  {author} {\bibinfo {author} {\bibfnamefont {E.}~\bibnamefont
  {Barausse}}, \bibinfo {author} {\bibfnamefont {C.}~\bibnamefont
  {Palenzuela}}, \bibinfo {author} {\bibfnamefont {M.}~\bibnamefont {Ponce}}, \
  and\ \bibinfo {author} {\bibfnamefont {L.}~\bibnamefont {Lehner}},\ }\href
  {\doibase 10.1103/PhysRevD.87.081506} {\bibfield  {journal} {\bibinfo
  {journal} {Phys. Rev. D}\ }\textbf {\bibinfo {volume} {87}},\ \bibinfo
  {pages} {081506(R)} (\bibinfo {year} {2013})},\ \Eprint
  {http://arxiv.org/abs/1212.5053} {arXiv:1212.5053 [gr-qc]} \BibitemShut
  {NoStop}%
\bibitem [{\citenamefont {Berti}\ \emph {et~al.}(2013)\citenamefont {Berti},
  \citenamefont {Cardoso}, \citenamefont {Gualtieri}, \citenamefont
  {Horbatsch},\ and\ \citenamefont {Sperhake}}]{Berti:2013gfa}%
  \BibitemOpen
  \bibfield  {author} {\bibinfo {author} {\bibfnamefont {E.}~\bibnamefont
  {Berti}}, \bibinfo {author} {\bibfnamefont {V.}~\bibnamefont {Cardoso}},
  \bibinfo {author} {\bibfnamefont {L.}~\bibnamefont {Gualtieri}}, \bibinfo
  {author} {\bibfnamefont {M.}~\bibnamefont {Horbatsch}}, \ and\ \bibinfo
  {author} {\bibfnamefont {U.}~\bibnamefont {Sperhake}},\ }\href {\doibase
  10.1103/PhysRevD.87.124020} {\bibfield  {journal} {\bibinfo  {journal} {Phys.
  Rev. D}\ }\textbf {\bibinfo {volume} {87}},\ \bibinfo {pages} {124020}
  (\bibinfo {year} {2013})},\ \Eprint {http://arxiv.org/abs/1304.2836}
  {arXiv:1304.2836 [gr-qc]} \BibitemShut {NoStop}%
\bibitem [{\citenamefont {Shibata}\ \emph {et~al.}(2014)\citenamefont
  {Shibata}, \citenamefont {Taniguchi}, \citenamefont {Okawa},\ and\
  \citenamefont {Buonanno}}]{Shibata:2013pra}%
  \BibitemOpen
  \bibfield  {author} {\bibinfo {author} {\bibfnamefont {M.}~\bibnamefont
  {Shibata}}, \bibinfo {author} {\bibfnamefont {K.}~\bibnamefont {Taniguchi}},
  \bibinfo {author} {\bibfnamefont {H.}~\bibnamefont {Okawa}}, \ and\ \bibinfo
  {author} {\bibfnamefont {A.}~\bibnamefont {Buonanno}},\ }\href {\doibase
  10.1103/PhysRevD.89.084005} {\bibfield  {journal} {\bibinfo  {journal} {Phys.
  Rev. D}\ }\textbf {\bibinfo {volume} {89}},\ \bibinfo {pages} {084005}
  (\bibinfo {year} {2014})},\ \Eprint {http://arxiv.org/abs/1310.0627}
  {arXiv:1310.0627 [gr-qc]} \BibitemShut {NoStop}%
\bibitem [{\citenamefont {Palenzuela}\ \emph {et~al.}(2014)\citenamefont
  {Palenzuela}, \citenamefont {Barausse}, \citenamefont {Ponce},\ and\
  \citenamefont {Lehner}}]{Palenzuela:2013hsa}%
  \BibitemOpen
  \bibfield  {author} {\bibinfo {author} {\bibfnamefont {C.}~\bibnamefont
  {Palenzuela}}, \bibinfo {author} {\bibfnamefont {E.}~\bibnamefont
  {Barausse}}, \bibinfo {author} {\bibfnamefont {M.}~\bibnamefont {Ponce}}, \
  and\ \bibinfo {author} {\bibfnamefont {L.}~\bibnamefont {Lehner}},\ }\href
  {\doibase 10.1103/PhysRevD.89.044024} {\bibfield  {journal} {\bibinfo
  {journal} {Phys. Rev. D}\ }\textbf {\bibinfo {volume} {89}},\ \bibinfo
  {pages} {044024} (\bibinfo {year} {2014})},\ \Eprint
  {http://arxiv.org/abs/1310.4481} {arXiv:1310.4481 [gr-qc]} \BibitemShut
  {NoStop}%
\bibitem [{\citenamefont {Taniguchi}\ \emph {et~al.}(2015)\citenamefont
  {Taniguchi}, \citenamefont {Shibata},\ and\ \citenamefont
  {Buonanno}}]{Taniguchi:2014fqa}%
  \BibitemOpen
  \bibfield  {author} {\bibinfo {author} {\bibfnamefont {K.}~\bibnamefont
  {Taniguchi}}, \bibinfo {author} {\bibfnamefont {M.}~\bibnamefont {Shibata}},
  \ and\ \bibinfo {author} {\bibfnamefont {A.}~\bibnamefont {Buonanno}},\
  }\href {\doibase 10.1103/PhysRevD.91.024033} {\bibfield  {journal} {\bibinfo
  {journal} {Phys. Rev. D}\ }\textbf {\bibinfo {volume} {91}},\ \bibinfo
  {pages} {024033} (\bibinfo {year} {2015})},\ \Eprint
  {http://arxiv.org/abs/1410.0738} {arXiv:1410.0738 [gr-qc]} \BibitemShut
  {NoStop}%
\bibitem [{\citenamefont {Damour}\ and\ \citenamefont
  {Esposito-Farese}(1992)}]{Damour:1992we}%
  \BibitemOpen
  \bibfield  {author} {\bibinfo {author} {\bibfnamefont {T.}~\bibnamefont
  {Damour}}\ and\ \bibinfo {author} {\bibfnamefont {G.}~\bibnamefont
  {Esposito-Farese}},\ }\href {\doibase 10.1088/0264-9381/9/9/015} {\bibfield
  {journal} {\bibinfo  {journal} {Class. Quant. Grav.}\ }\textbf {\bibinfo
  {volume} {9}},\ \bibinfo {pages} {2093} (\bibinfo {year} {1992})}\BibitemShut
  {NoStop}%
\bibitem [{\citenamefont {Damour}\ and\ \citenamefont
  {Esposito-Farese}(1993)}]{Damour:1993hw}%
  \BibitemOpen
  \bibfield  {author} {\bibinfo {author} {\bibfnamefont {T.}~\bibnamefont
  {Damour}}\ and\ \bibinfo {author} {\bibfnamefont {G.}~\bibnamefont
  {Esposito-Farese}},\ }\href {\doibase 10.1103/PhysRevLett.70.2220} {\bibfield
   {journal} {\bibinfo  {journal} {Phys. Rev. Lett.}\ }\textbf {\bibinfo
  {volume} {70}},\ \bibinfo {pages} {2220} (\bibinfo {year}
  {1993})}\BibitemShut {NoStop}%
\bibitem [{\citenamefont {Mirshekari}\ and\ \citenamefont
  {Will}(2013)}]{Mirshekari:2013vb}%
  \BibitemOpen
  \bibfield  {author} {\bibinfo {author} {\bibfnamefont {S.}~\bibnamefont
  {Mirshekari}}\ and\ \bibinfo {author} {\bibfnamefont {C.~M.}\ \bibnamefont
  {Will}},\ }\href {\doibase 10.1103/PhysRevD.87.084070} {\bibfield  {journal}
  {\bibinfo  {journal} {Phys. Rev. D}\ }\textbf {\bibinfo {volume} {87}},\
  \bibinfo {pages} {084070} (\bibinfo {year} {2013})},\ \Eprint
  {http://arxiv.org/abs/1301.4680} {arXiv:1301.4680 [gr-qc]} \BibitemShut
  {NoStop}%
\bibitem [{\citenamefont {Zaglauer}(1992)}]{Zaglauer:1992bp}%
  \BibitemOpen
  \bibfield  {author} {\bibinfo {author} {\bibfnamefont {H.~W.}\ \bibnamefont
  {Zaglauer}},\ }\href {\doibase 10.1086/171537} {\bibfield  {journal}
  {\bibinfo  {journal} {Astrophys. J.}\ }\textbf {\bibinfo {volume} {393}},\
  \bibinfo {pages} {685} (\bibinfo {year} {1992})}\BibitemShut {NoStop}%
\bibitem [{\citenamefont {Will}(2014)}]{Will:2014kxa}%
  \BibitemOpen
  \bibfield  {author} {\bibinfo {author} {\bibfnamefont {C.~M.}\ \bibnamefont
  {Will}},\ }\href {\doibase 10.12942/lrr-2014-4} {\bibfield  {journal}
  {\bibinfo  {journal} {Living Rev. Rel.}\ }\textbf {\bibinfo {volume} {17}},\
  \bibinfo {pages} {4} (\bibinfo {year} {2014})},\ \Eprint
  {http://arxiv.org/abs/1403.7377} {arXiv:1403.7377 [gr-qc]} \BibitemShut
  {NoStop}%
\bibitem [{\citenamefont {Lang}(2014)}]{Lang:2013fna}%
  \BibitemOpen
  \bibfield  {author} {\bibinfo {author} {\bibfnamefont {R.~N.}\ \bibnamefont
  {Lang}},\ }\href {\doibase 10.1103/PhysRevD.89.084014} {\bibfield  {journal}
  {\bibinfo  {journal} {Phys. Rev. D}\ }\textbf {\bibinfo {volume} {89}},\
  \bibinfo {pages} {084014} (\bibinfo {year} {2014})},\ \Eprint
  {http://arxiv.org/abs/1310.3320} {arXiv:1310.3320 [gr-qc]} \BibitemShut
  {NoStop}%
\bibitem [{\citenamefont {Lang}(2015)}]{Lang:2014osa}%
  \BibitemOpen
  \bibfield  {author} {\bibinfo {author} {\bibfnamefont {R.~N.}\ \bibnamefont
  {Lang}},\ }\href {\doibase 10.1103/PhysRevD.91.084027} {\bibfield  {journal}
  {\bibinfo  {journal} {Phys. Rev. D}\ }\textbf {\bibinfo {volume} {91}},\
  \bibinfo {pages} {084027} (\bibinfo {year} {2015})},\ \Eprint
  {http://arxiv.org/abs/1411.3073} {arXiv:1411.3073 [gr-qc]} \BibitemShut
  {NoStop}%
\bibitem [{\citenamefont {Bernard}(2018)}]{Bernard:2018hta}%
  \BibitemOpen
  \bibfield  {author} {\bibinfo {author} {\bibfnamefont {L.}~\bibnamefont
  {Bernard}},\ }\href {\doibase 10.1103/PhysRevD.98.044004} {\bibfield
  {journal} {\bibinfo  {journal} {Phys. Rev. D}\ }\textbf {\bibinfo {volume}
  {98}},\ \bibinfo {pages} {044004} (\bibinfo {year} {2018})},\ \Eprint
  {http://arxiv.org/abs/1802.10201} {arXiv:1802.10201 [gr-qc]} \BibitemShut
  {NoStop}%
\bibitem [{\citenamefont {Bernard}(2019)}]{Bernard:2018ivi}%
  \BibitemOpen
  \bibfield  {author} {\bibinfo {author} {\bibfnamefont {L.}~\bibnamefont
  {Bernard}},\ }\href {\doibase 10.1103/PhysRevD.99.044047} {\bibfield
  {journal} {\bibinfo  {journal} {Phys. Rev. D}\ }\textbf {\bibinfo {volume}
  {99}},\ \bibinfo {pages} {044047} (\bibinfo {year} {2019})},\ \Eprint
  {http://arxiv.org/abs/1812.04169} {arXiv:1812.04169 [gr-qc]} \BibitemShut
  {NoStop}%
\bibitem [{\citenamefont {Brax}\ \emph {et~al.}(2021)\citenamefont {Brax},
  \citenamefont {Davis}, \citenamefont {Melville},\ and\ \citenamefont
  {Wong}}]{Brax:2021qqo}%
  \BibitemOpen
  \bibfield  {author} {\bibinfo {author} {\bibfnamefont {P.}~\bibnamefont
  {Brax}}, \bibinfo {author} {\bibfnamefont {A.-C.}\ \bibnamefont {Davis}},
  \bibinfo {author} {\bibfnamefont {S.}~\bibnamefont {Melville}}, \ and\
  \bibinfo {author} {\bibfnamefont {L.~K.}\ \bibnamefont {Wong}},\ }\href
  {\doibase 10.1088/1475-7516/2021/10/075} {\bibfield  {journal} {\bibinfo
  {journal} {JCAP}\ }\textbf {\bibinfo {volume} {10}},\ \bibinfo {pages} {075}
  (\bibinfo {year} {2021})},\ \Eprint {http://arxiv.org/abs/2107.10841}
  {arXiv:2107.10841 [gr-qc]} \BibitemShut {NoStop}%
\bibitem [{\citenamefont {Shiralilou}\ \emph {et~al.}(2022)\citenamefont
  {Shiralilou}, \citenamefont {Hinderer}, \citenamefont {Nissanke},
  \citenamefont {Ortiz},\ and\ \citenamefont {Witek}}]{Shiralilou:2021mfl}%
  \BibitemOpen
  \bibfield  {author} {\bibinfo {author} {\bibfnamefont {B.}~\bibnamefont
  {Shiralilou}}, \bibinfo {author} {\bibfnamefont {T.}~\bibnamefont
  {Hinderer}}, \bibinfo {author} {\bibfnamefont {S.~M.}\ \bibnamefont
  {Nissanke}}, \bibinfo {author} {\bibfnamefont {N.}~\bibnamefont {Ortiz}}, \
  and\ \bibinfo {author} {\bibfnamefont {H.}~\bibnamefont {Witek}},\ }\href
  {\doibase 10.1088/1361-6382/ac4196} {\bibfield  {journal} {\bibinfo
  {journal} {Class. Quant. Grav.}\ }\textbf {\bibinfo {volume} {39}},\ \bibinfo
  {pages} {035002} (\bibinfo {year} {2022})},\ \Eprint
  {http://arxiv.org/abs/2105.13972} {arXiv:2105.13972 [gr-qc]} \BibitemShut
  {NoStop}%
\bibitem [{\citenamefont {Khalil}\ \emph {et~al.}(2022)\citenamefont {Khalil},
  \citenamefont {Mendes}, \citenamefont {Ortiz},\ and\ \citenamefont
  {Steinhoff}}]{Khalil:2022sii}%
  \BibitemOpen
  \bibfield  {author} {\bibinfo {author} {\bibfnamefont {M.}~\bibnamefont
  {Khalil}}, \bibinfo {author} {\bibfnamefont {R.~F.~P.}\ \bibnamefont
  {Mendes}}, \bibinfo {author} {\bibfnamefont {N.}~\bibnamefont {Ortiz}}, \
  and\ \bibinfo {author} {\bibfnamefont {J.}~\bibnamefont {Steinhoff}},\ }\href
  {\doibase 10.1103/PhysRevD.106.104016} {\bibfield  {journal} {\bibinfo
  {journal} {Phys. Rev. D}\ }\textbf {\bibinfo {volume} {106}},\ \bibinfo
  {pages} {104016} (\bibinfo {year} {2022})},\ \Eprint
  {http://arxiv.org/abs/2206.13233} {arXiv:2206.13233 [gr-qc]} \BibitemShut
  {NoStop}%
\bibitem [{\citenamefont {Bernard}\ \emph {et~al.}(2022)\citenamefont
  {Bernard}, \citenamefont {Blanchet},\ and\ \citenamefont
  {Trestini}}]{Bernard:2022noq}%
  \BibitemOpen
  \bibfield  {author} {\bibinfo {author} {\bibfnamefont {L.}~\bibnamefont
  {Bernard}}, \bibinfo {author} {\bibfnamefont {L.}~\bibnamefont {Blanchet}}, \
  and\ \bibinfo {author} {\bibfnamefont {D.}~\bibnamefont {Trestini}},\ }\href
  {\doibase 10.1088/1475-7516/2022/08/008} {\bibfield  {journal} {\bibinfo
  {journal} {JCAP}\ }\textbf {\bibinfo {volume} {08}},\ \bibinfo {pages} {008}
  (\bibinfo {year} {2022})},\ \Eprint {http://arxiv.org/abs/2201.10924}
  {arXiv:2201.10924 [gr-qc]} \BibitemShut {NoStop}%
\bibitem [{\citenamefont {Bernard}\ \emph {et~al.}(2024)\citenamefont
  {Bernard}, \citenamefont {Dones},\ and\ \citenamefont
  {Mougiakakos}}]{Bernard:2023eul}%
  \BibitemOpen
  \bibfield  {author} {\bibinfo {author} {\bibfnamefont {L.}~\bibnamefont
  {Bernard}}, \bibinfo {author} {\bibfnamefont {E.}~\bibnamefont {Dones}}, \
  and\ \bibinfo {author} {\bibfnamefont {S.}~\bibnamefont {Mougiakakos}},\
  }\href {\doibase 10.1103/PhysRevD.109.044006} {\bibfield  {journal} {\bibinfo
   {journal} {Phys. Rev. D}\ }\textbf {\bibinfo {volume} {109}},\ \bibinfo
  {pages} {044006} (\bibinfo {year} {2024})},\ \Eprint
  {http://arxiv.org/abs/2310.19679} {arXiv:2310.19679 [gr-qc]} \BibitemShut
  {NoStop}%
\bibitem [{\citenamefont {Chowdhuri}\ and\ \citenamefont
  {Bhattacharyya}(2022)}]{AbhishekChowdhuri:2022ora}%
  \BibitemOpen
  \bibfield  {author} {\bibinfo {author} {\bibfnamefont {A.}~\bibnamefont
  {Chowdhuri}}\ and\ \bibinfo {author} {\bibfnamefont {A.}~\bibnamefont
  {Bhattacharyya}},\ }\href {\doibase 10.1103/PhysRevD.106.064046} {\bibfield
  {journal} {\bibinfo  {journal} {Phys. Rev. D}\ }\textbf {\bibinfo {volume}
  {106}},\ \bibinfo {pages} {064046} (\bibinfo {year} {2022})},\ \Eprint
  {http://arxiv.org/abs/2203.09917} {arXiv:2203.09917 [gr-qc]} \BibitemShut
  {NoStop}%
\bibitem [{\citenamefont {Jain}\ and\ \citenamefont
  {Rettegno}(2024)}]{Jain:2024lie}%
  \BibitemOpen
  \bibfield  {author} {\bibinfo {author} {\bibfnamefont {T.}~\bibnamefont
  {Jain}}\ and\ \bibinfo {author} {\bibfnamefont {P.}~\bibnamefont
  {Rettegno}},\ }\href@noop {} {\  (\bibinfo {year} {2024})},\ \Eprint
  {http://arxiv.org/abs/2407.10908} {arXiv:2407.10908 [gr-qc]} \BibitemShut
  {NoStop}%
\bibitem [{\citenamefont {Juli\'e}\ and\ \citenamefont
  {Deruelle}(2017)}]{Julie:2017pkb}%
  \BibitemOpen
  \bibfield  {author} {\bibinfo {author} {\bibfnamefont {F.-L.}\ \bibnamefont
  {Juli\'e}}\ and\ \bibinfo {author} {\bibfnamefont {N.}~\bibnamefont
  {Deruelle}},\ }\href {\doibase 10.1103/PhysRevD.95.124054} {\bibfield
  {journal} {\bibinfo  {journal} {Phys. Rev. D}\ }\textbf {\bibinfo {volume}
  {95}},\ \bibinfo {pages} {124054} (\bibinfo {year} {2017})},\ \Eprint
  {http://arxiv.org/abs/1703.05360} {arXiv:1703.05360 [gr-qc]} \BibitemShut
  {NoStop}%
\bibitem [{\citenamefont {Juli\'e}(2018)}]{Julie:2017ucp}%
  \BibitemOpen
  \bibfield  {author} {\bibinfo {author} {\bibfnamefont {F.-L.}\ \bibnamefont
  {Juli\'e}},\ }\href {\doibase 10.1103/PhysRevD.97.024047} {\bibfield
  {journal} {\bibinfo  {journal} {Phys. Rev. D}\ }\textbf {\bibinfo {volume}
  {97}},\ \bibinfo {pages} {024047} (\bibinfo {year} {2018})},\ \Eprint
  {http://arxiv.org/abs/1709.09742} {arXiv:1709.09742 [gr-qc]} \BibitemShut
  {NoStop}%
\bibitem [{\citenamefont {Jain}\ \emph {et~al.}(2023)\citenamefont {Jain},
  \citenamefont {Rettegno}, \citenamefont {Agathos}, \citenamefont {Nagar},\
  and\ \citenamefont {Turco}}]{Jain:2022nxs}%
  \BibitemOpen
  \bibfield  {author} {\bibinfo {author} {\bibfnamefont {T.}~\bibnamefont
  {Jain}}, \bibinfo {author} {\bibfnamefont {P.}~\bibnamefont {Rettegno}},
  \bibinfo {author} {\bibfnamefont {M.}~\bibnamefont {Agathos}}, \bibinfo
  {author} {\bibfnamefont {A.}~\bibnamefont {Nagar}}, \ and\ \bibinfo {author}
  {\bibfnamefont {L.}~\bibnamefont {Turco}},\ }\href {\doibase
  10.1103/PhysRevD.107.084017} {\bibfield  {journal} {\bibinfo  {journal}
  {Phys. Rev. D}\ }\textbf {\bibinfo {volume} {107}},\ \bibinfo {pages}
  {084017} (\bibinfo {year} {2023})},\ \Eprint
  {http://arxiv.org/abs/2211.15580} {arXiv:2211.15580 [gr-qc]} \BibitemShut
  {NoStop}%
\bibitem [{\citenamefont {Jain}(2023)}]{Jain:2023fvt}%
  \BibitemOpen
  \bibfield  {author} {\bibinfo {author} {\bibfnamefont {T.}~\bibnamefont
  {Jain}},\ }\href {\doibase 10.1103/PhysRevD.107.084018} {\bibfield  {journal}
  {\bibinfo  {journal} {Phys. Rev. D}\ }\textbf {\bibinfo {volume} {107}},\
  \bibinfo {pages} {084018} (\bibinfo {year} {2023})},\ \Eprint
  {http://arxiv.org/abs/2301.01070} {arXiv:2301.01070 [gr-qc]} \BibitemShut
  {NoStop}%
\bibitem [{\citenamefont {Juli\'e}\ \emph {et~al.}(2023)\citenamefont
  {Juli\'e}, \citenamefont {Baibhav}, \citenamefont {Berti},\ and\
  \citenamefont {Buonanno}}]{Julie:2022qux}%
  \BibitemOpen
  \bibfield  {author} {\bibinfo {author} {\bibfnamefont {F.-L.}\ \bibnamefont
  {Juli\'e}}, \bibinfo {author} {\bibfnamefont {V.}~\bibnamefont {Baibhav}},
  \bibinfo {author} {\bibfnamefont {E.}~\bibnamefont {Berti}}, \ and\ \bibinfo
  {author} {\bibfnamefont {A.}~\bibnamefont {Buonanno}},\ }\href {\doibase
  10.1103/PhysRevD.107.104044} {\bibfield  {journal} {\bibinfo  {journal}
  {Phys. Rev. D}\ }\textbf {\bibinfo {volume} {107}},\ \bibinfo {pages}
  {104044} (\bibinfo {year} {2023})},\ \Eprint
  {http://arxiv.org/abs/2212.13802} {arXiv:2212.13802 [gr-qc]} \BibitemShut
  {NoStop}%
\bibitem [{\citenamefont {Buonanno}\ and\ \citenamefont
  {Damour}(1999)}]{Buonanno:1998gg}%
  \BibitemOpen
  \bibfield  {author} {\bibinfo {author} {\bibfnamefont {A.}~\bibnamefont
  {Buonanno}}\ and\ \bibinfo {author} {\bibfnamefont {T.}~\bibnamefont
  {Damour}},\ }\href {\doibase 10.1103/PhysRevD.59.084006} {\bibfield
  {journal} {\bibinfo  {journal} {Phys. Rev.}\ }\textbf {\bibinfo {volume}
  {D59}},\ \bibinfo {pages} {084006} (\bibinfo {year} {1999})},\ \Eprint
  {http://arxiv.org/abs/gr-qc/9811091} {arXiv:gr-qc/9811091} \BibitemShut
  {NoStop}%
\bibitem [{\citenamefont {Buonanno}\ and\ \citenamefont
  {Damour}(2000)}]{Buonanno:2000ef}%
  \BibitemOpen
  \bibfield  {author} {\bibinfo {author} {\bibfnamefont {A.}~\bibnamefont
  {Buonanno}}\ and\ \bibinfo {author} {\bibfnamefont {T.}~\bibnamefont
  {Damour}},\ }\href {\doibase 10.1103/PhysRevD.62.064015} {\bibfield
  {journal} {\bibinfo  {journal} {Phys. Rev.}\ }\textbf {\bibinfo {volume}
  {D62}},\ \bibinfo {pages} {064015} (\bibinfo {year} {2000})},\ \Eprint
  {http://arxiv.org/abs/gr-qc/0001013} {arXiv:gr-qc/0001013} \BibitemShut
  {NoStop}%
\bibitem [{\citenamefont {Iyer}\ and\ \citenamefont
  {Will}(1993)}]{Iyer:1993xi}%
  \BibitemOpen
  \bibfield  {author} {\bibinfo {author} {\bibfnamefont {B.~R.}\ \bibnamefont
  {Iyer}}\ and\ \bibinfo {author} {\bibfnamefont {C.~M.}\ \bibnamefont
  {Will}},\ }\href {\doibase 10.1103/PhysRevLett.70.113} {\bibfield  {journal}
  {\bibinfo  {journal} {Phys. Rev. Lett.}\ }\textbf {\bibinfo {volume} {70}},\
  \bibinfo {pages} {113} (\bibinfo {year} {1993})}\BibitemShut {NoStop}%
\bibitem [{\citenamefont {Iyer}\ and\ \citenamefont
  {Will}(1995)}]{Iyer:1995rn}%
  \BibitemOpen
  \bibfield  {author} {\bibinfo {author} {\bibfnamefont {B.~R.}\ \bibnamefont
  {Iyer}}\ and\ \bibinfo {author} {\bibfnamefont {C.~M.}\ \bibnamefont
  {Will}},\ }\href {\doibase 10.1103/PhysRevD.52.6882} {\bibfield  {journal}
  {\bibinfo  {journal} {Phys. Rev. D}\ }\textbf {\bibinfo {volume} {52}},\
  \bibinfo {pages} {6882} (\bibinfo {year} {1995})}\BibitemShut {NoStop}%
\bibitem [{\citenamefont {Damour}\ and\ \citenamefont
  {Esposito-Farese}(1996{\natexlab{b}})}]{Damour:1995kt}%
  \BibitemOpen
  \bibfield  {author} {\bibinfo {author} {\bibfnamefont {T.}~\bibnamefont
  {Damour}}\ and\ \bibinfo {author} {\bibfnamefont {G.}~\bibnamefont
  {Esposito-Farese}},\ }\href {\doibase 10.1103/PhysRevD.53.5541} {\bibfield
  {journal} {\bibinfo  {journal} {Phys. Rev. D}\ }\textbf {\bibinfo {volume}
  {53}},\ \bibinfo {pages} {5541} (\bibinfo {year} {1996}{\natexlab{b}})},\
  \Eprint {http://arxiv.org/abs/gr-qc/9506063} {arXiv:gr-qc/9506063}
  \BibitemShut {NoStop}%
\bibitem [{\citenamefont {{Eardley}}(1975)}]{1975ApJ...196L..59E}%
  \BibitemOpen
  \bibfield  {author} {\bibinfo {author} {\bibfnamefont {D.~M.}\ \bibnamefont
  {{Eardley}}},\ }\href {\doibase 10.1086/181744} {\bibfield  {journal}
  {\bibinfo  {journal} {apjl}\ }\textbf {\bibinfo {volume} {196}},\ \bibinfo
  {pages} {L59} (\bibinfo {year} {1975})}\BibitemShut {NoStop}%
\bibitem [{\citenamefont {Lange}\ \emph {et~al.}(2017)\citenamefont {Lange}
  \emph {et~al.}}]{Lange:2017wki}%
  \BibitemOpen
  \bibfield  {author} {\bibinfo {author} {\bibfnamefont {J.}~\bibnamefont
  {Lange}} \emph {et~al.},\ }\href {\doibase 10.1103/PhysRevD.96.104041}
  {\bibfield  {journal} {\bibinfo  {journal} {Phys. Rev. D}\ }\textbf {\bibinfo
  {volume} {96}},\ \bibinfo {pages} {104041} (\bibinfo {year} {2017})},\
  \Eprint {http://arxiv.org/abs/1705.09833} {arXiv:1705.09833 [gr-qc]}
  \BibitemShut {NoStop}%
\bibitem [{\citenamefont {Lange}\ \emph {et~al.}(2018)\citenamefont {Lange},
  \citenamefont {O'Shaughnessy},\ and\ \citenamefont {Rizzo}}]{Lange:2018pyp}%
  \BibitemOpen
  \bibfield  {author} {\bibinfo {author} {\bibfnamefont {J.}~\bibnamefont
  {Lange}}, \bibinfo {author} {\bibfnamefont {R.}~\bibnamefont
  {O'Shaughnessy}}, \ and\ \bibinfo {author} {\bibfnamefont {M.}~\bibnamefont
  {Rizzo}},\ }\href@noop {} {\  (\bibinfo {year} {2018})},\ \Eprint
  {http://arxiv.org/abs/1805.10457} {arXiv:1805.10457 [gr-qc]} \BibitemShut
  {NoStop}%
\bibitem [{\citenamefont {Damour}\ \emph {et~al.}(2000)\citenamefont {Damour},
  \citenamefont {Iyer},\ and\ \citenamefont {Sathyaprakash}}]{Damour:2000gg}%
  \BibitemOpen
  \bibfield  {author} {\bibinfo {author} {\bibfnamefont {T.}~\bibnamefont
  {Damour}}, \bibinfo {author} {\bibfnamefont {B.~R.}\ \bibnamefont {Iyer}}, \
  and\ \bibinfo {author} {\bibfnamefont {B.}~\bibnamefont {Sathyaprakash}},\
  }\href {\doibase 10.1103/PhysRevD.62.084036} {\bibfield  {journal} {\bibinfo
  {journal} {Phys.Rev.}\ }\textbf {\bibinfo {volume} {D62}},\ \bibinfo {pages}
  {084036} (\bibinfo {year} {2000})},\ \Eprint
  {http://arxiv.org/abs/gr-qc/0001023} {arXiv:gr-qc/0001023 [gr-qc]}
  \BibitemShut {NoStop}%
\bibitem [{\citenamefont {Bini}\ and\ \citenamefont
  {Damour}(2012)}]{Bini:2012ji}%
  \BibitemOpen
  \bibfield  {author} {\bibinfo {author} {\bibfnamefont {D.}~\bibnamefont
  {Bini}}\ and\ \bibinfo {author} {\bibfnamefont {T.}~\bibnamefont {Damour}},\
  }\href {\doibase 10.1103/PhysRevD.86.124012} {\bibfield  {journal} {\bibinfo
  {journal} {Phys.Rev.}\ }\textbf {\bibinfo {volume} {D86}},\ \bibinfo {pages}
  {124012} (\bibinfo {year} {2012})},\ \Eprint {http://arxiv.org/abs/1210.2834}
  {arXiv:1210.2834 [gr-qc]} \BibitemShut {NoStop}%
\bibitem [{\citenamefont {Khalil}\ \emph {et~al.}(2021)\citenamefont {Khalil},
  \citenamefont {Buonanno}, \citenamefont {Steinhoff},\ and\ \citenamefont
  {Vines}}]{Khalil:2021txt}%
  \BibitemOpen
  \bibfield  {author} {\bibinfo {author} {\bibfnamefont {M.}~\bibnamefont
  {Khalil}}, \bibinfo {author} {\bibfnamefont {A.}~\bibnamefont {Buonanno}},
  \bibinfo {author} {\bibfnamefont {J.}~\bibnamefont {Steinhoff}}, \ and\
  \bibinfo {author} {\bibfnamefont {J.}~\bibnamefont {Vines}},\ }\href
  {\doibase 10.1103/PhysRevD.104.024046} {\bibfield  {journal} {\bibinfo
  {journal} {Phys. Rev. D}\ }\textbf {\bibinfo {volume} {104}},\ \bibinfo
  {pages} {024046} (\bibinfo {year} {2021})},\ \Eprint
  {http://arxiv.org/abs/2104.11705} {arXiv:2104.11705 [gr-qc]} \BibitemShut
  {NoStop}%
\bibitem [{\citenamefont {Buonanno}\ \emph {et~al.}(2006)\citenamefont
  {Buonanno}, \citenamefont {Chen},\ and\ \citenamefont
  {Damour}}]{Buonanno:2005xu}%
  \BibitemOpen
  \bibfield  {author} {\bibinfo {author} {\bibfnamefont {A.}~\bibnamefont
  {Buonanno}}, \bibinfo {author} {\bibfnamefont {Y.}~\bibnamefont {Chen}}, \
  and\ \bibinfo {author} {\bibfnamefont {T.}~\bibnamefont {Damour}},\ }\href
  {\doibase 10.1103/PhysRevD.74.104005} {\bibfield  {journal} {\bibinfo
  {journal} {Phys. Rev.}\ }\textbf {\bibinfo {volume} {D74}},\ \bibinfo {pages}
  {104005} (\bibinfo {year} {2006})},\ \Eprint
  {http://arxiv.org/abs/gr-qc/0508067} {arXiv:gr-qc/0508067} \BibitemShut
  {NoStop}%
\bibitem [{\citenamefont {Schott}(1915)}]{Schott}%
  \BibitemOpen
  \bibfield  {author} {\bibinfo {author} {\bibfnamefont {G.}~\bibnamefont
  {Schott}},\ }\href {\doibase 10.1080/14786440108635280} {\bibfield  {journal}
  {\bibinfo  {journal} {The London, Edinburgh, and Dublin Philosophical
  Magazine and Journal of Science}\ }\textbf {\bibinfo {volume} {29}},\
  \bibinfo {pages} {49} (\bibinfo {year} {1915})},\ \Eprint
  {http://arxiv.org/abs/https://doi.org/10.1080/14786440108635280}
  {https://doi.org/10.1080/14786440108635280} \BibitemShut {NoStop}%
\bibitem [{\citenamefont {Landau}\ and\ \citenamefont
  {Lifshitz}(1975)}]{landau1975classical}%
  \BibitemOpen
  \bibfield  {author} {\bibinfo {author} {\bibfnamefont {L.~D.}\ \bibnamefont
  {Landau}}\ and\ \bibinfo {author} {\bibfnamefont {E.~M.}\ \bibnamefont
  {Lifshitz}},\ }\href@noop {} {\emph {\bibinfo {title} {The Classical Theory
  of Fields}}},\ \bibinfo {edition} {4th}\ ed.,\ \bibinfo {series} {Course of
  Theoretical Physics}, Vol.~\bibinfo {volume} {2}\ (\bibinfo  {publisher}
  {Pergamon Press},\ \bibinfo {address} {Oxford},\ \bibinfo {year} {1975})\
  \bibinfo {note} {section 75}\BibitemShut {NoStop}%
\bibitem [{\citenamefont {Hinderer}\ and\ \citenamefont
  {Babak}(2017)}]{Hinderer:2017jcs}%
  \BibitemOpen
  \bibfield  {author} {\bibinfo {author} {\bibfnamefont {T.}~\bibnamefont
  {Hinderer}}\ and\ \bibinfo {author} {\bibfnamefont {S.}~\bibnamefont
  {Babak}},\ }\href {\doibase 10.1103/PhysRevD.96.104048} {\bibfield  {journal}
  {\bibinfo  {journal} {Phys. Rev.}\ }\textbf {\bibinfo {volume} {D96}},\
  \bibinfo {pages} {104048} (\bibinfo {year} {2017})},\ \Eprint
  {http://arxiv.org/abs/1707.08426} {arXiv:1707.08426 [gr-qc]} \BibitemShut
  {NoStop}%
\bibitem [{\citenamefont {Gamboa}\ \emph {et~al.}(2024)\citenamefont {Gamboa}
  \emph {et~al.}}]{Gamboa:2024hli}%
  \BibitemOpen
  \bibfield  {author} {\bibinfo {author} {\bibfnamefont {A.}~\bibnamefont
  {Gamboa}} \emph {et~al.},\ }\href@noop {} {\  (\bibinfo {year} {2024})},\
  \Eprint {http://arxiv.org/abs/2412.12823} {arXiv:2412.12823 [gr-qc]}
  \BibitemShut {NoStop}%
\bibitem [{\citenamefont {Albanesi}\ \emph {et~al.}(2025)\citenamefont
  {Albanesi}, \citenamefont {Gamba}, \citenamefont {Bernuzzi}, \citenamefont
  {Fontbut{\'e}}, \citenamefont {Gonzalez},\ and\ \citenamefont
  {Nagar}}]{Albanesi:2025txj}%
  \BibitemOpen
  \bibfield  {author} {\bibinfo {author} {\bibfnamefont {S.}~\bibnamefont
  {Albanesi}}, \bibinfo {author} {\bibfnamefont {R.}~\bibnamefont {Gamba}},
  \bibinfo {author} {\bibfnamefont {S.}~\bibnamefont {Bernuzzi}}, \bibinfo
  {author} {\bibfnamefont {J.}~\bibnamefont {Fontbut{\'e}}}, \bibinfo {author}
  {\bibfnamefont {A.}~\bibnamefont {Gonzalez}}, \ and\ \bibinfo {author}
  {\bibfnamefont {A.}~\bibnamefont {Nagar}},\ }\href@noop {} {\  (\bibinfo
  {year} {2025})},\ \Eprint {http://arxiv.org/abs/2503.14580} {arXiv:2503.14580
  [gr-qc]} \BibitemShut {NoStop}%
\end{thebibliography}%
	
\end{document}